\documentclass[lettersize,journal]{IEEEtran}
\usepackage{amsmath,amsfonts}
\usepackage{algorithmic}
\usepackage{algorithm}
\usepackage{array}
\usepackage[caption=false,font=scriptsize,labelfont=sf,textfont=sf]{subfig}
\usepackage{textcomp}
\usepackage{stfloats}
\usepackage{url}
\usepackage{verbatim}
\usepackage{graphicx}
\usepackage{cite}
\usepackage{multirow}
\usepackage{flushend}

\def\BibTeX{{\rm B\kern-.05em{\sc i\kern-.025em b}\kern-.08em
    T\kern-.1667em\lower.7ex\hbox{E}\kern-.125emX}}
\usepackage{balance}
\begin{document}
\title{Classification Method of Road Surface Condition and Type with LiDAR Using Spatiotemporal Information}
\author{Ju Won Seo
\IEEEmembership{Student Member, IEEE}, Jin Sung Kim \IEEEmembership{Graduate Student Member, IEEE}, Jin Ho Yang \IEEEmembership{Graduate Student Member, IEEE}, and Chung Choo Chung \IEEEmembership{Member, IEEE}
\thanks{This work was supported by the National
Research Foundation of Korea (NRF) through the Korea Government Ministry
of Science and Information and Communication Technology (MSIT), Data Driven Optimized Autonomous Driving Technology Using Open Set Classification Method, under Grant 2021R1A2C2009908.}
\thanks{Ju Won Seo, Jin Sung Kim, and Jin Ho Yang are with Dept. of Electrical Engineeing, Hanyang University, Seoul 04763, Korea (email : \{suhju1227, jskim06, jjz0426\}@hanyang.ac.kr)}
\thanks{C. C. Chung is with Div. of Electrical and Biomedical Engineering, Hanyang University, Seoul 04763, Korea (email: cchung@hanyang.ac.kr)}
}

\markboth{xxx,~VOL.~xx, NO.~xx, xxxx~2023}%
{Seo \MakeLowercase{\textit{et al.}}: Classification method of road surface condition and type with LiDAR using spatiotemporal information}

\maketitle

\begin{abstract}
This paper proposes a spatiotemporal architecture with a deep neural network  (DNN) for road surface conditions and types classification using LiDAR. It is known that LiDAR provides information on the reflectivity and number of point clouds depending on a road surface. Thus, this paper utilizes the information to classify the road surface. We divided the front road area into four subregions. First, we constructed feature vectors using each subregion's reflectivity, number of point clouds, and in-vehicle information. Second, the DNN classifies road surface conditions and types for each subregion. Finally, the output of the DNN feeds into the spatiotemporal process to make the final classification reflecting vehicle speed and probability given by the outcomes of softmax functions of the DNN output layer. To validate the effectiveness of the proposed method, we performed a comparative study with five other algorithms. With the proposed DNN, we obtained the highest accuracy of 98.0\% and 98.6\% for two subregions near the vehicle. In addition, we implemented the proposed method on the Jetson TX2 board to confirm that it is applicable in real-time.
\end{abstract}


\section{Introduction}
\IEEEPARstart{I}{n} automated driving, accurate detection of road surface conditions and types are essential because the friction between the road surface and the tire has consequences for driver safety~\cite{wallman2001friction}. Hydroplaning by heavy rain or low friction between tires and slippery roads influenced by heavy snow can cause fatal car accidents. In addition, knowing the types of roads provide essential information regarding safety, fuel efficiency, and driver convenience. Determining the types of roads also estimates various physical quantities, including friction coefficients, slip angles, and vehicle handling characteristics~\cite{brooks2005vibration}. Thus, many techniques for classifying the conditions and types of road surfaces have been studied.

Various sensors have been used to classify road conditions and types.
Generally, cameras have been widely used to detect road surface conditions and types. One technique is to classify the road surface conditions using a camera sensor with convolutional neural networks (CNN), which performs well in image classification~\cite{cheng2019road,nolte2018assessment}. In addition, feature extraction of camera image using gray-level co-occurrence matrix (GLCM) was used to classify the type of road surface~\cite{marianingsih2018comparison}.
Another approach to determine road surface conditions is to use in-vehicle sensors~\cite{berntorp2018tire, kim2019comparative, kim2018interacting}. Through a neural network (NN), the road conditions are determined using the acceleration data collected by the in-vehicle sensors with a fast Fourier transform (FFT)~\cite{kim2020road}. In addition, the standard sensor in commercialized vehicles was utilized to obtain the road surface condition using the fuzzy logic block and artificial NN~\cite{castillo2015robust}. Despite various studies using these sensors, road surface classification requires another approach. Using cameras with illumination is particularly challenging, and in-vehicle sensors may have harsh noise. Therefore research has been conducted using another sensor to compensate for these problems.

LiDAR can be effective in determining road surface conditions and types. Using LiDAR is robust against various factors, such as the shadow of objects and illumination changes. It is known that LiDAR provides information on the reflectivity and range depending on the roughness of objects. Using these characteristics of LiDAR, many studies have applied them to classifying road surfaces.
One study classified road conditions using regular reflection on the wet road due to water film and irregular reflection on dry roads~\cite{aki2016road}.
Likewise, there is a study to classify the types of roads by utilizing the differences in the range and remission of LiDAR data since the roughness varies depending on the type of road ~\cite{wang2012terrain}. Various studies have been conducted to classify road surfaces using the characteristics of LiDAR that can be obtained according to road surfaces.

In general, support vector machine (SVM) and K-nearest neighbor (KNN) techniques have been utilized to classify the condition and type of road surface using LiDAR~\cite{wang2012terrain,walas2014terrain,heinzler2019weather}. After extracting the reflectivity and range of LiDAR point cloud as a feature, methods of classifying the types of roads through SVM have been studied~\cite{wang2012terrain,walas2014terrain}. Furthermore, KNN is used to classify the road condition using reflectivity information from LiDAR~\cite{heinzler2019weather}. Another approach is to transform LiDAR data into an image. In~\cite{sebastian2021range}, the authors proposed a method of determining the road condition and weather with a convolutional encoder by transforming the LiDAR point cloud into a range image. These studies determined the condition and type of roads using only the characteristics of LiDAR, such as reflectivity and range. However, it is necessary to utilize the spatiotemporal information of sensor data to classify the road surface in an autonomous vehicle since the feature of LiDAR varies as the vehicle moves~\cite{merriaux2017movingcar}.

This paper proposes a spatiotemporal architecture with the LiDAR and in-vehicle sensor data for classifying road conditions and types. To get spatial information, we divide the front road area into four subregions: left near region, right near region, left far region, and right far region.
We classify the road conditions and types for each region using a DNN. The features are reflectivity, number of point clouds, and ego vehicle speed. The selected features are stacked using the time-windowing method to utilize the temporal information, and this feature vector constructs the input vector of DNN. Among the four divided subregions, we focus on the regions near the vehicle since they are notably more important than the road further away by considering the braking distance for the driver's safety. Therefore, we construct a spatiotemporal formulation with past DNN results in regions far from the vehicle, calculating the final results of regions close to the vehicle.
We compared the results with five other methods (SVM, KNN, etc.) to validate the proposed algorithm. The experimental results show that the proposed method outperforms the other methods by obtaining the accuracies, $98.0\%$ and $98.6\%$, for each near left and right subregion, respectively. Furthermore, we implemented the proposed method on the Jetson TX2 board to confirm that it is applicable in real-time. It was similar to the results implemented on a desktop ,and the average computation time was about 10 ms.

\section{Approach for Road Surface Condition and Type Classification}
\label{sec:approach}
\subsection{Scenario Description}
\label{sebsec:scenario}
\begin{figure}[t]
\centering
\includegraphics[width=0.45\textwidth]{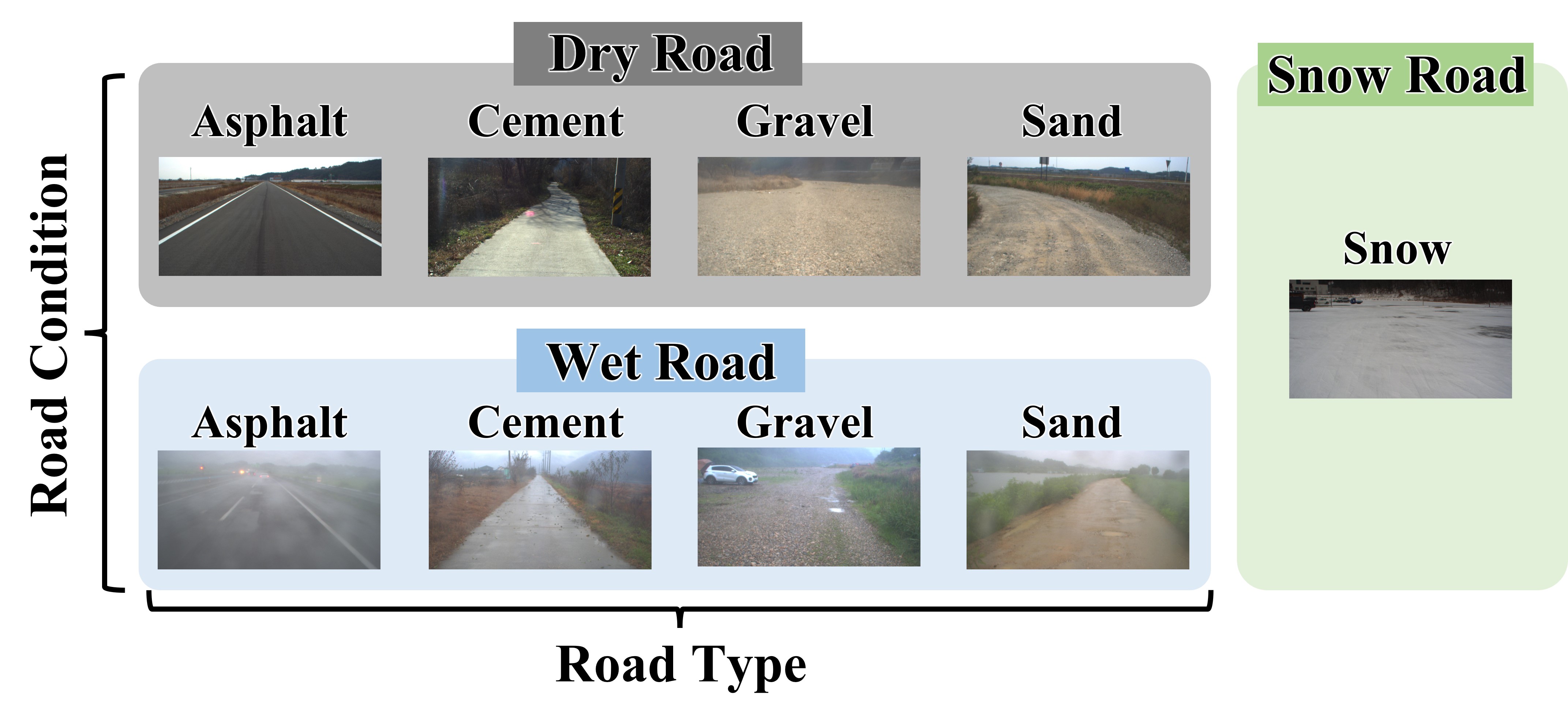}
\caption{Road conditions (dry, wet, and snow) and types (asphalt, cement, gravel, and sand) with obtained sample images.}
\label{fig:conditions_types}
\end{figure}
This study considers three road conditions (dry, wet, and snow) and four road types (asphalt, cement, gravel, and sand). Figure~\ref{fig:conditions_types} shows the obtained sample images of each road condition and type. In an urban area, it is common to discover that different types of roads exist, and rain or snow may affect the condition of the road surface. Therefore, classifying the condition and type of road surface is a big issue in autonomous vehicles in urban areas. Thus, this paper considers the driving environment in an urban area with various conditions and types of road surfaces. We assumed the Car-to-Car Rear braking (CCRb) scenario of autonomous emergency braking (AEB) protocol from the Euro NCAP so that the ego vehicle speed is limited to under $50$ km/h~\cite{schram2013implementation}.

\subsection{Data Acquisition}
\label{subsec:DataAcquisition}

\begin{figure}[t]
\centering
\subfloat[][Experiment vehicle]{\includegraphics[width=0.35\textwidth]{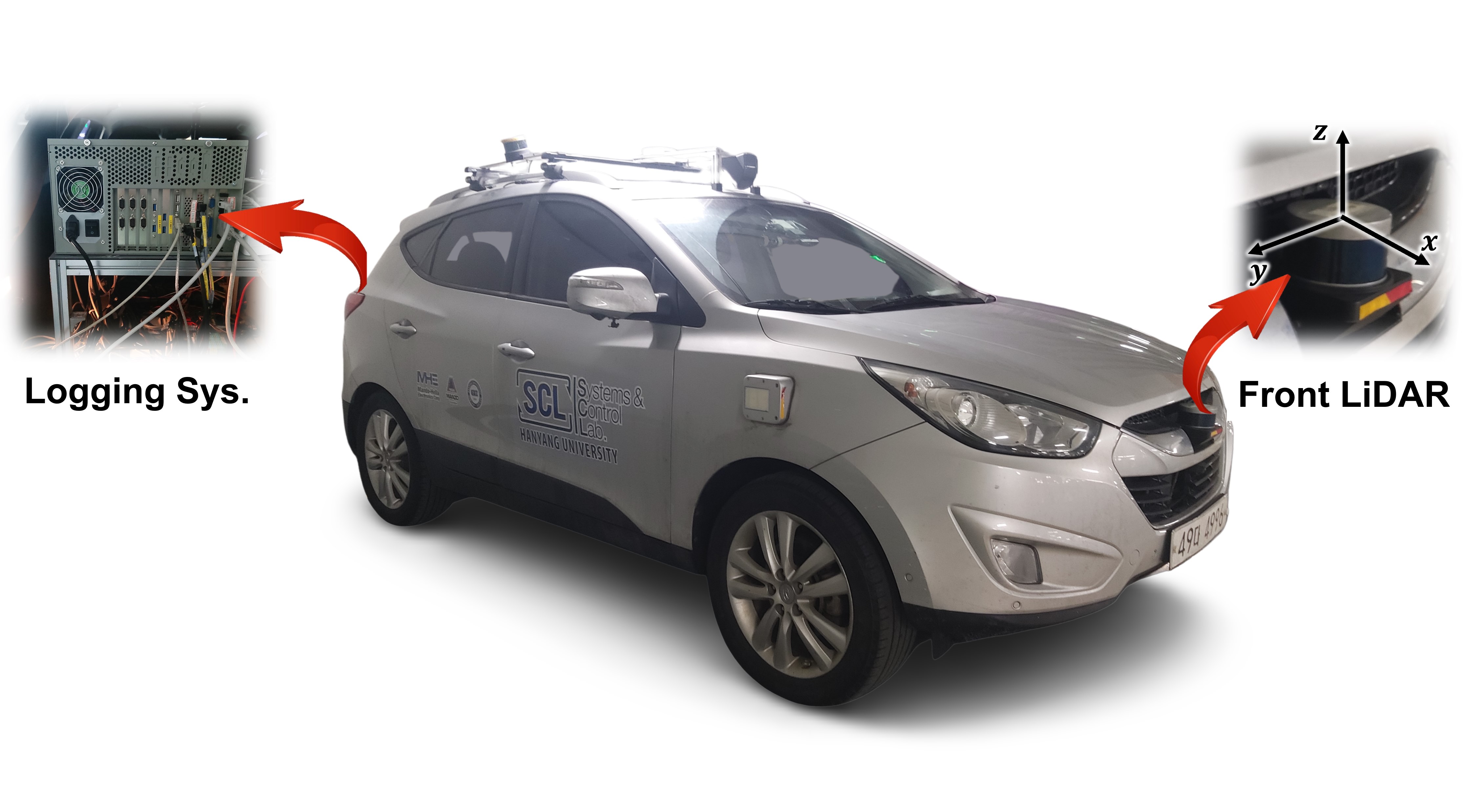}}
\hfil
\subfloat[][Sensor logging system]{\includegraphics[width=0.35\textwidth]{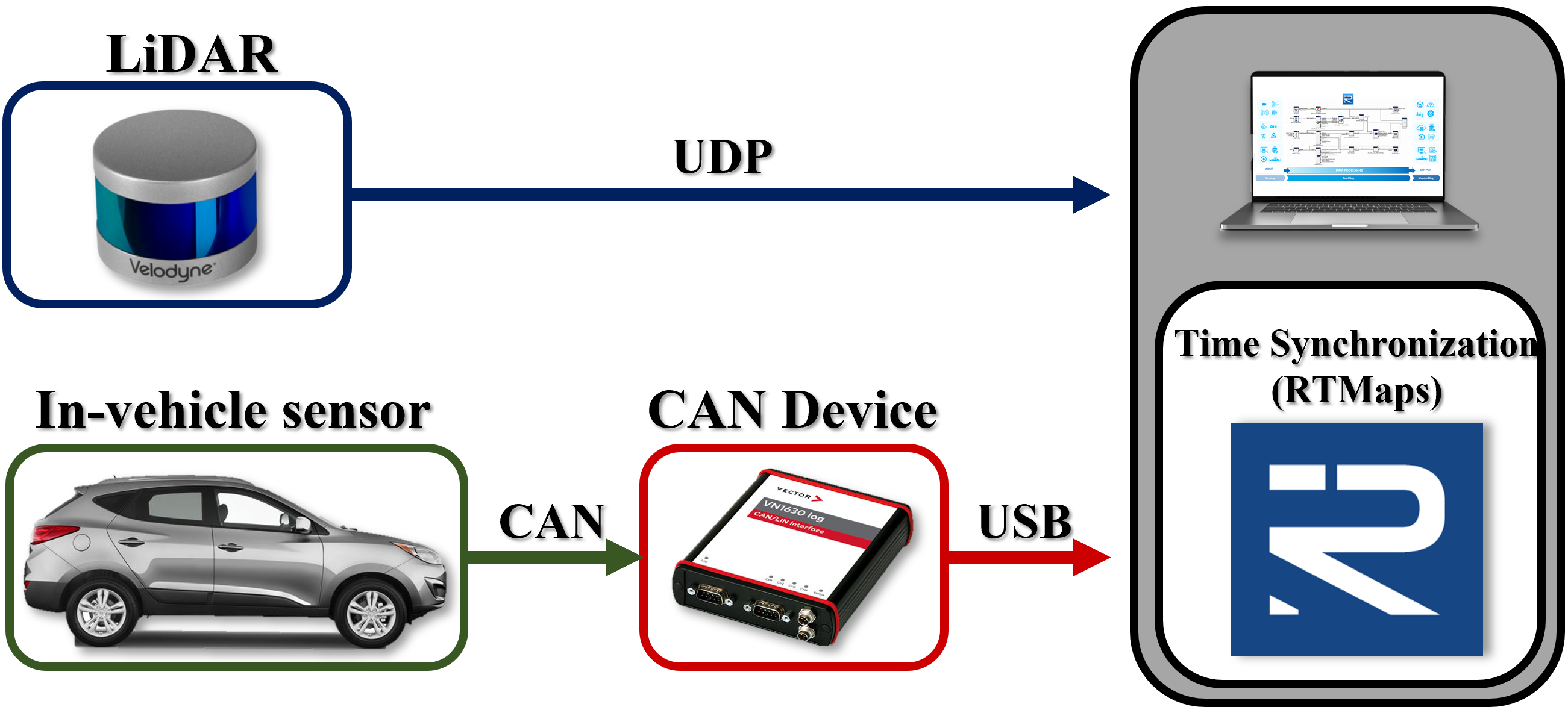}}
\caption{{Configuration of the logging system in our test vehicle. We use the test vehicle with LiDAR and in-vehicle sensors for obtaining the data to make the training data set}}
\label{fig:vehicle_sensor}
\end{figure}

In this study, we use a real car to obtain the dataset. Figure~\ref{fig:vehicle_sensor} shows the experimental vehicle and sensor system used for the data acquisition. The vehicle is a small SUV, Tucson ix from Hyundai Motors, and sensor data was obtained from the LiDAR and in-vehicle sensors. The LiDAR was mounted 0.85 m above the ground in the vehicle's front bumper so that the LiDAR could obtain data on the front road surface.
The VLP16 from Velodyne, which has a rotation rate of $10~Hz$ and 16 vertical channels, is used.
As shown in Fig.~\ref{fig:vehicle_sensor}~(b), the VN1630 from Vector is adopted to obtain the data from the in-vehicle sensors through the controller area network (CAN) bus. The vehicle state data was obtained from the in-vehicle sensors, e.g., the vehicle's speed. A logging system is needed to obtain synchronized data from various communication methods. Here, the RTMaps from Intempora was utilized for acquiring LiDAR and in-vehicle data under time synchronization.

The dataset was obtained for three road conditions (dry, wet, and snow) and four road types (asphalt, cement, gravel, and sand) where the vehicle drives in various places. The point cloud obtained when LiDAR rotates once with a rotation rate of 10 Hz is regarded as one frame. Point cloud data of 34,000 frames from LiDAR and vehicle speed data from the in-vehicle sensor were obtained for each road class, so a total of about 9 hours of point cloud data and vehicle speed data were obtained. After generating a feature vector from the obtained data (will be explained in Section~\ref{subsec:FeatureVector}), 243,000 frames were used as a training dataset. The remaining 63,000 frames were used as a validation dataset, as shown in Table~\ref{tb:data}. The validation dataset consists of data acquired from different places on different days from the training data set.

\begin{table}[b]
\centering
\caption{Data set for training and validation obtained by the logging system}
\label{tb:data}
\begin{tabular}{|cc|c|c|}
\hline
\multicolumn{2}{|c|}{\textbf{Road}}                                    & \textbf{Training Dataset}                                    & \textbf{Validation Dataset}                               \\ \hline
\multicolumn{1}{|c|}{\multirow{4}{*}{\textbf{Dry}}} & \textbf{Asphalt} & \begin{tabular}[c]{@{}c@{}}27,000\\ (2,700 sec)\end{tabular} & \begin{tabular}[c]{@{}c@{}}7,000\\ (700 sec)\end{tabular} \\ \cline{2-4}
\multicolumn{1}{|c|}{}                              & \textbf{Cement}  & \begin{tabular}[c]{@{}c@{}}27,000\\ (2,700 sec)\end{tabular} & \begin{tabular}[c]{@{}c@{}}7,000\\ (700 sec)\end{tabular} \\ \cline{2-4}
\multicolumn{1}{|c|}{}                              & \textbf{Gravel}  & \begin{tabular}[c]{@{}c@{}}27,000\\ (2,700 sec)\end{tabular} & \begin{tabular}[c]{@{}c@{}}7,000\\ (700 sec)\end{tabular} \\ \cline{2-4}
\multicolumn{1}{|c|}{}                              & \textbf{Sand}    & \begin{tabular}[c]{@{}c@{}}27,000\\ (2,700 sec)\end{tabular} & \begin{tabular}[c]{@{}c@{}}7,000\\ (700 sec)\end{tabular} \\ \hline
\multicolumn{1}{|c|}{\multirow{4}{*}{\textbf{Wet}}} & \textbf{Asphalt} & \begin{tabular}[c]{@{}c@{}}27,000\\ (2,700 sec)\end{tabular} & \begin{tabular}[c]{@{}c@{}}7,000\\ (700 sec)\end{tabular} \\ \cline{2-4}
\multicolumn{1}{|c|}{}                              & \textbf{Cement}  & \begin{tabular}[c]{@{}c@{}}27,000\\ (2,700 sec)\end{tabular} & \begin{tabular}[c]{@{}c@{}}7,000\\ (700 sec)\end{tabular} \\ \cline{2-4}
\multicolumn{1}{|c|}{}                              & \textbf{Gravel}  & \begin{tabular}[c]{@{}c@{}}27,000\\ (2,700 sec)\end{tabular} & \begin{tabular}[c]{@{}c@{}}7,000\\ (700 sec)\end{tabular} \\ \cline{2-4}
\multicolumn{1}{|c|}{}                              & \textbf{Sand}    & \begin{tabular}[c]{@{}c@{}}27,000\\ (2,700 sec)\end{tabular} & \begin{tabular}[c]{@{}c@{}}7,000\\ (700 sec)\end{tabular} \\ \hline
\multicolumn{2}{|c|}{\textbf{Snow}}                                    & \begin{tabular}[c]{@{}c@{}}27,000\\ (2,700 sec)\end{tabular} & \begin{tabular}[c]{@{}c@{}}7,000\\ (700 sec)\end{tabular} \\ \hline
\end{tabular}
\end{table}
\subsection{Region Tessellation}
\label{subsec:SectionDivide}

\begin{figure}[t]
\centering
\includegraphics[width=0.4\textwidth]{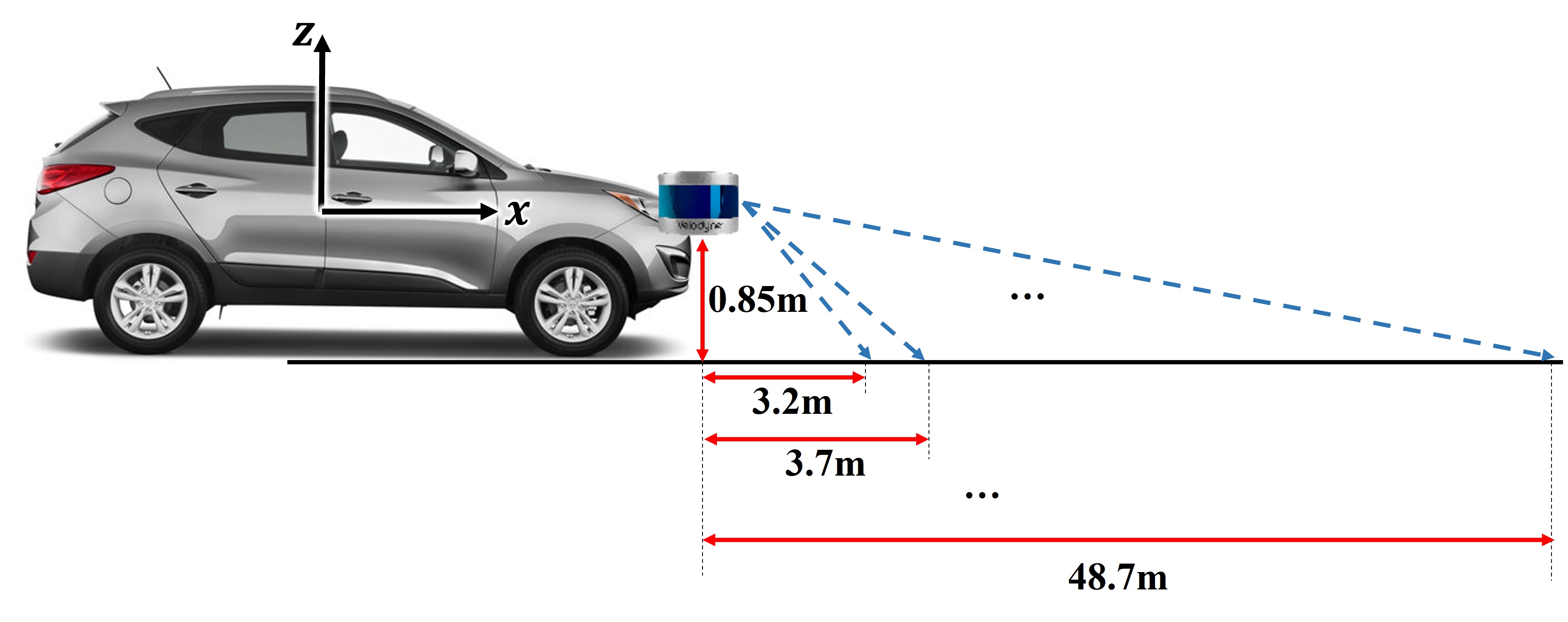}
\caption{Detecting points at each LiDAR channel, which measures the road surface. The nearest distance is about 3.2m and the farthest distance is about 48.7m.}
\label{fig:LiDAR_distance}
\end{figure}
\begin{figure}[t]
\centering
\hspace{-0.5cm}
{\includegraphics[width=0.4\textwidth]{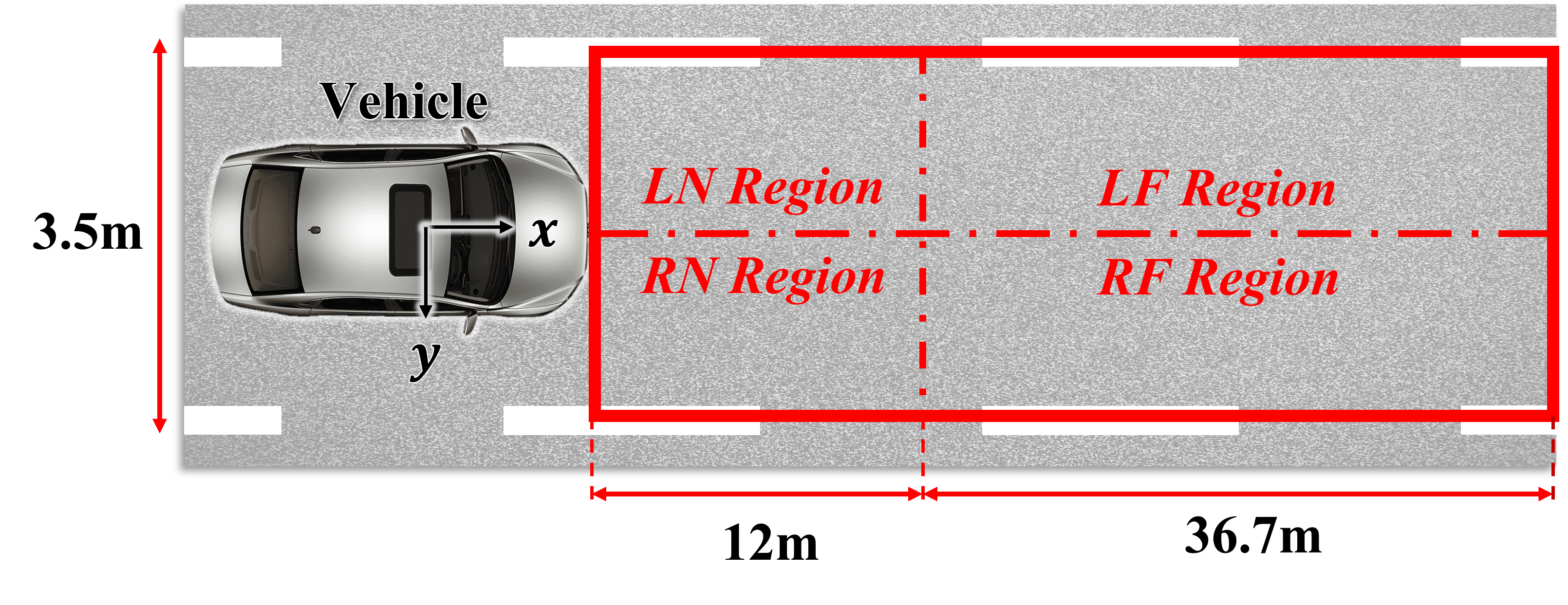}}
\caption{{Divided regions of the front road according to the AEB test protocol and road regulations. Since the regions close to the vehicle is essential regarding the safety of the vehicle, \emph{LN Region} and \emph{RN Region} are mainly considered in this paper.}}
\label{fig:Section}
\end{figure}

This paper proposes an algorithm that classifies road conditions and types with spatiotemporal information. To obtain this information, we first need to set the region of interest (ROI) of the LiDAR sensor. We used the data in which the LiDAR detected the front road surface. Hence, the data was selected by the z-axis criteria below $0.1$ m.
In addition, the lateral distance was set from $-1.75$ m to $1.75$ m according to the available road width of 3.5 m. The longitudinal distance was designed as $48.7$ m, the longest detectable distance for LiDAR, as depicted in Fig~\ref{fig:LiDAR_distance}.
For spatial information, we divided the ROI of the front road into four subregions, as shown in Fig.~\ref{fig:Section}. Here, \emph{L} means \emph{Left}, \emph{R} means \emph{Right}, \emph{N} means \emph{Near the vehicle}, and \emph{F} means \emph{Far from the vehicle}. For example, the \emph{LF Region} means the \emph{Region of the left and far from the vehicle}.
In the lateral direction, the ROI was separated as the \{\emph{LN Region}, \emph{LF Region}\} group and the \{\emph{RN Region}, \emph{RF Region}\} group. This is because the road conditions can significantly differ between the left and right wheel path, the so-called split friction. The split friction phenomenon can affect the safety of the vehicle's motion. Thus, we separated the ROI into the left and right regions to classify each wheel path's road conditions.
Moreover, we divided the ROI into the near and far regions as the \{\emph{LN Region}, \emph{RN Region}\} group and the \{\emph{LF Region}, \emph{RF Region}\} group.
According to the Car-to-Car Rear braking (CCRb) scenario of AEB test protocol from Euro NCAP~\cite{schram2013implementation},
the near region in the longitudinal direction was set as $12$ m to guarantee collision avoidance with the front car. In this paper, \emph{N Regions} are mainly considered because the conditions of the road close to the vehicle are notably more important than the road further away.
\subsection{Feature Vector Selection}
\label{subsec:FeatureVector}
\begin{figure*}[t]
\centering
\subfloat[][LF Region]{\includegraphics[width=0.239\textwidth]{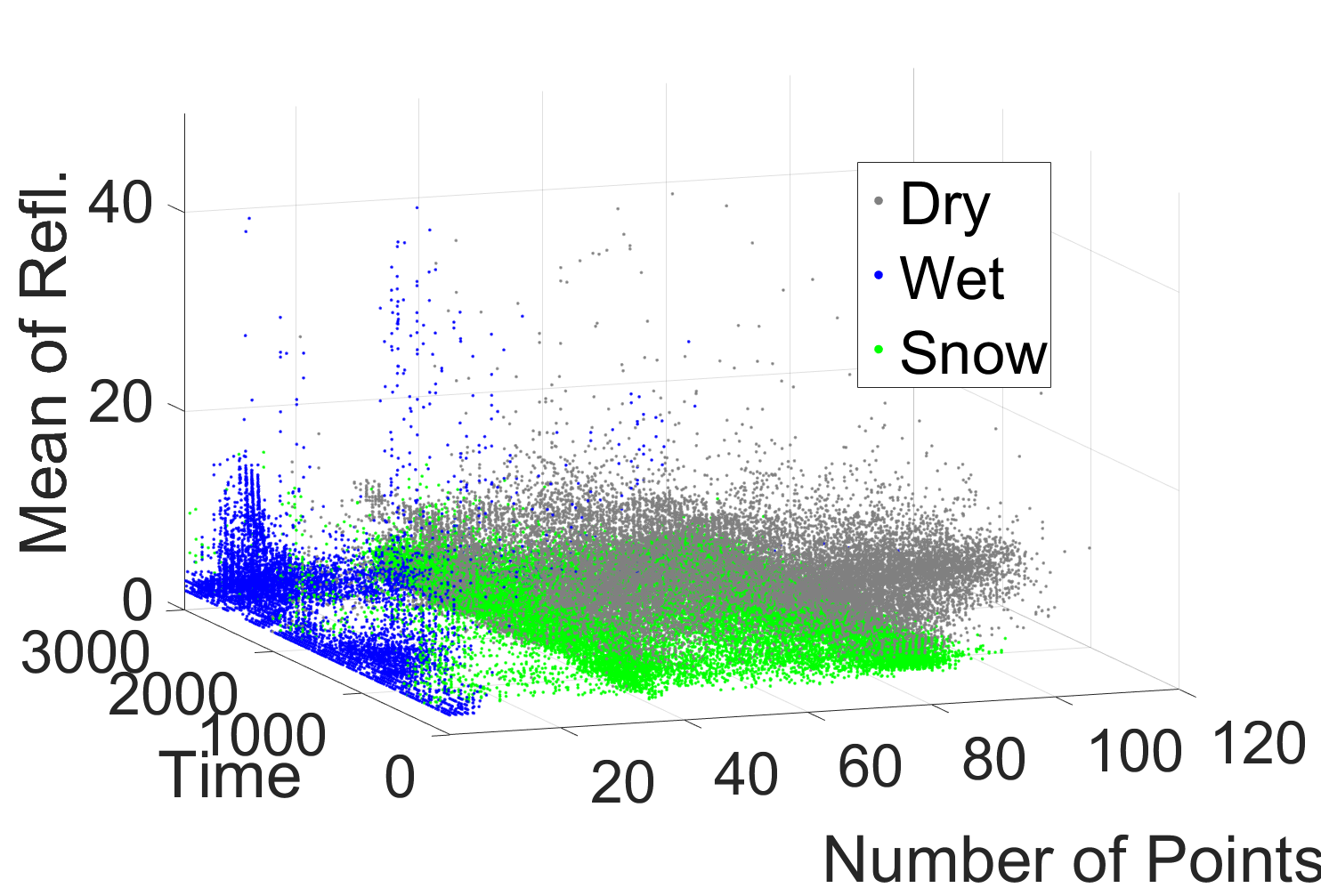}}
\subfloat[][RF Region]{\includegraphics[width=0.239\textwidth]{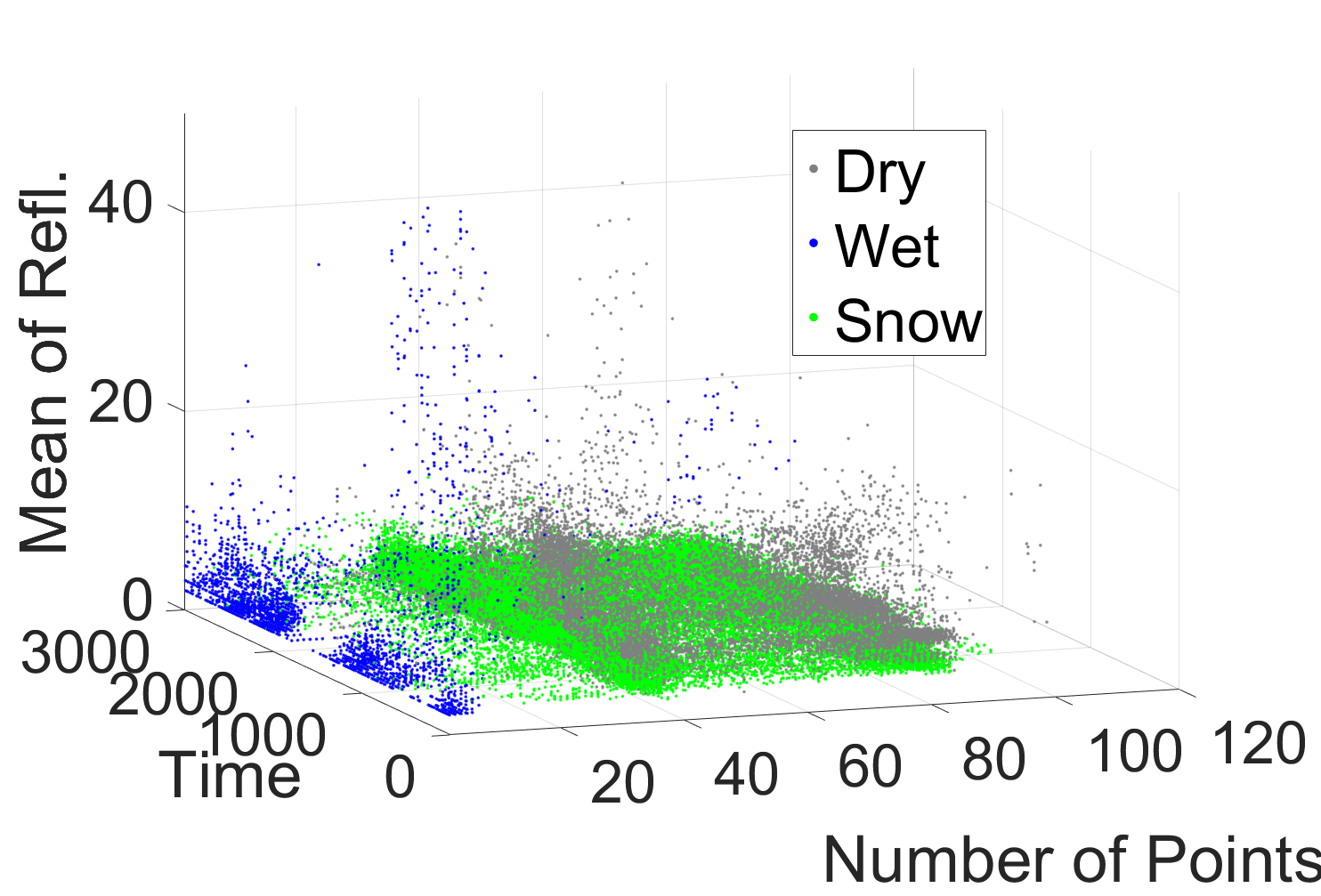}}
\subfloat[][LN Region]{\includegraphics[width=0.239\textwidth]{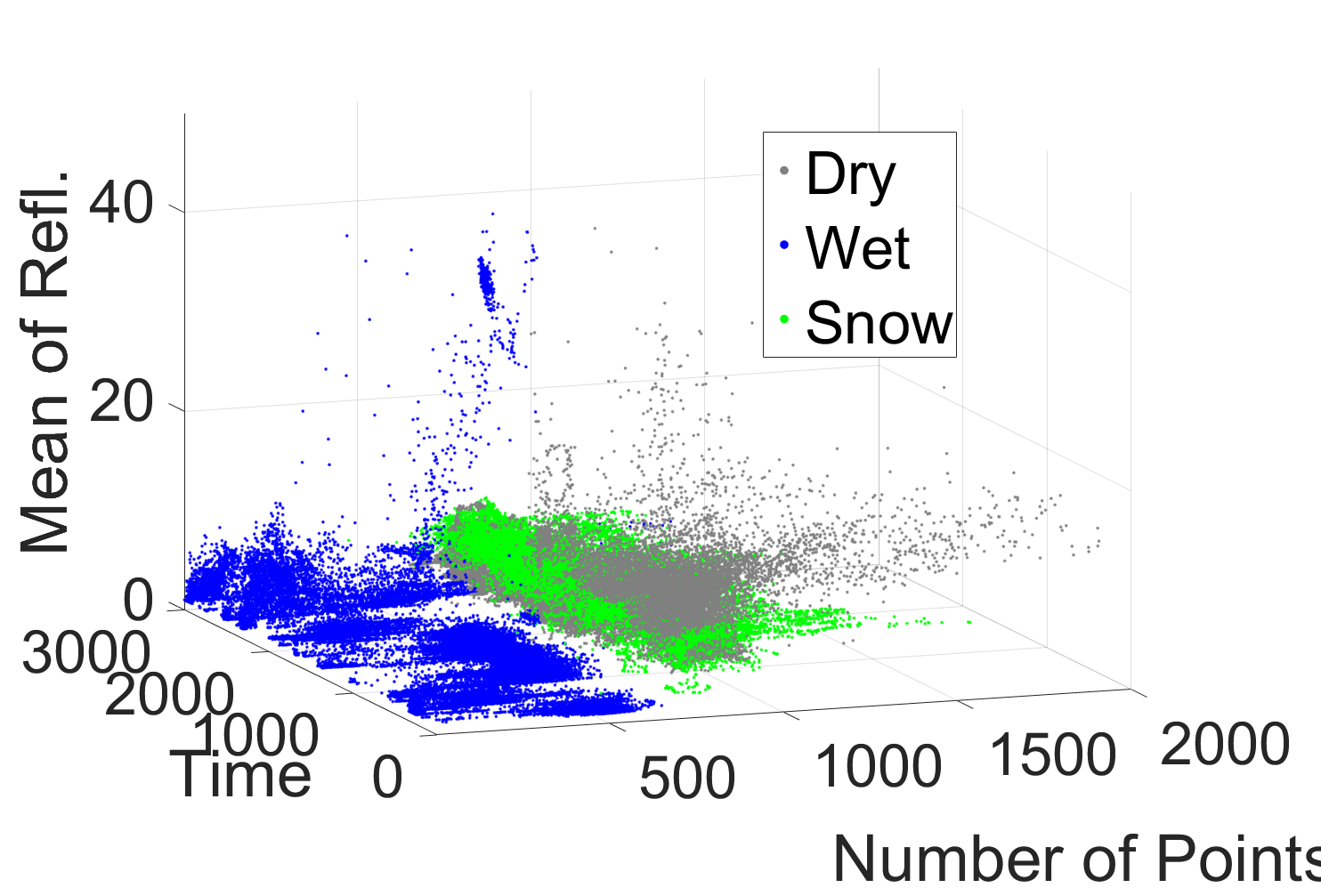}}
\subfloat[][RN Region]{\includegraphics[width=0.239\textwidth]{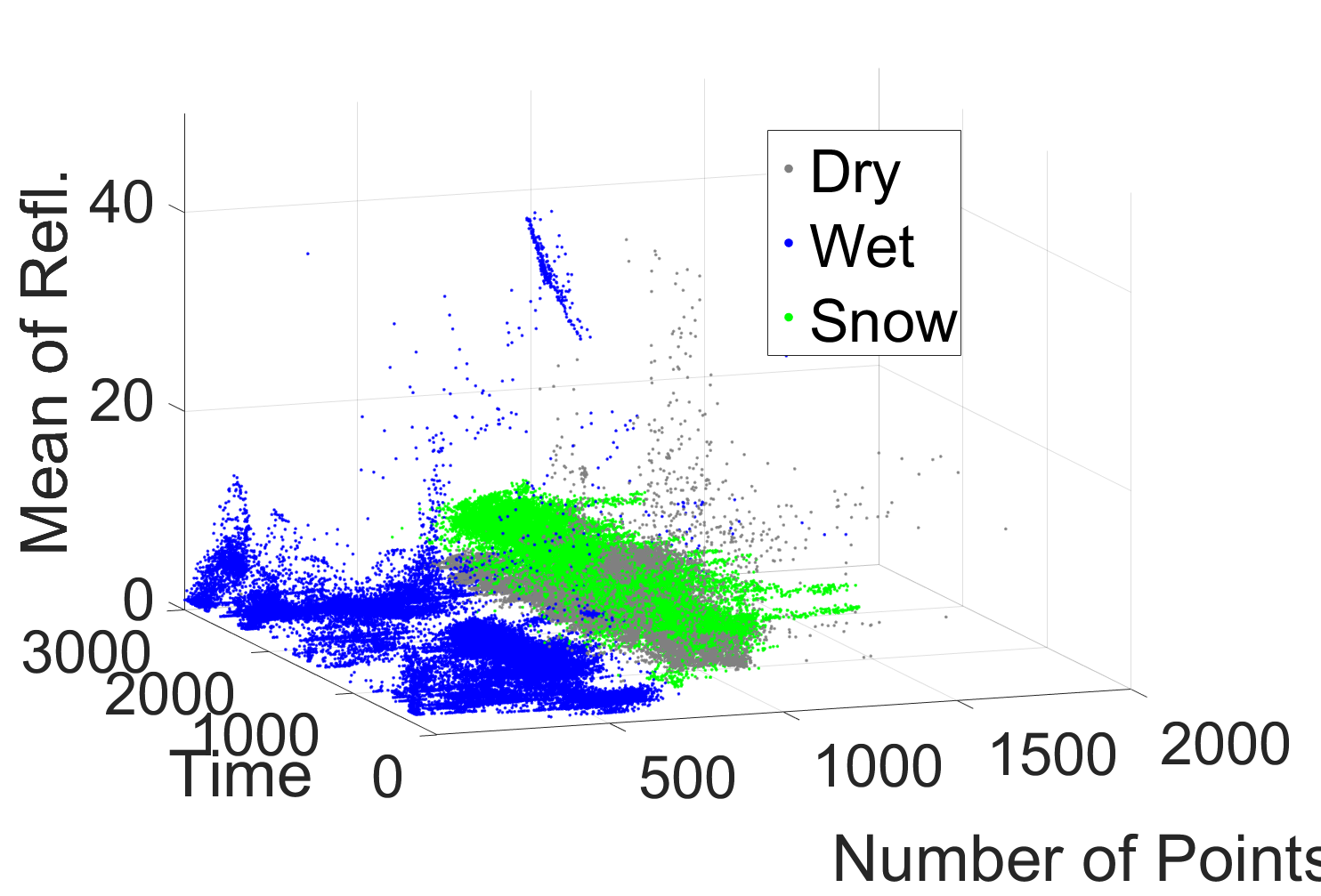}}
\hfil
\subfloat[][LF Region Num.-Refl.]{\includegraphics[width=0.239\textwidth]{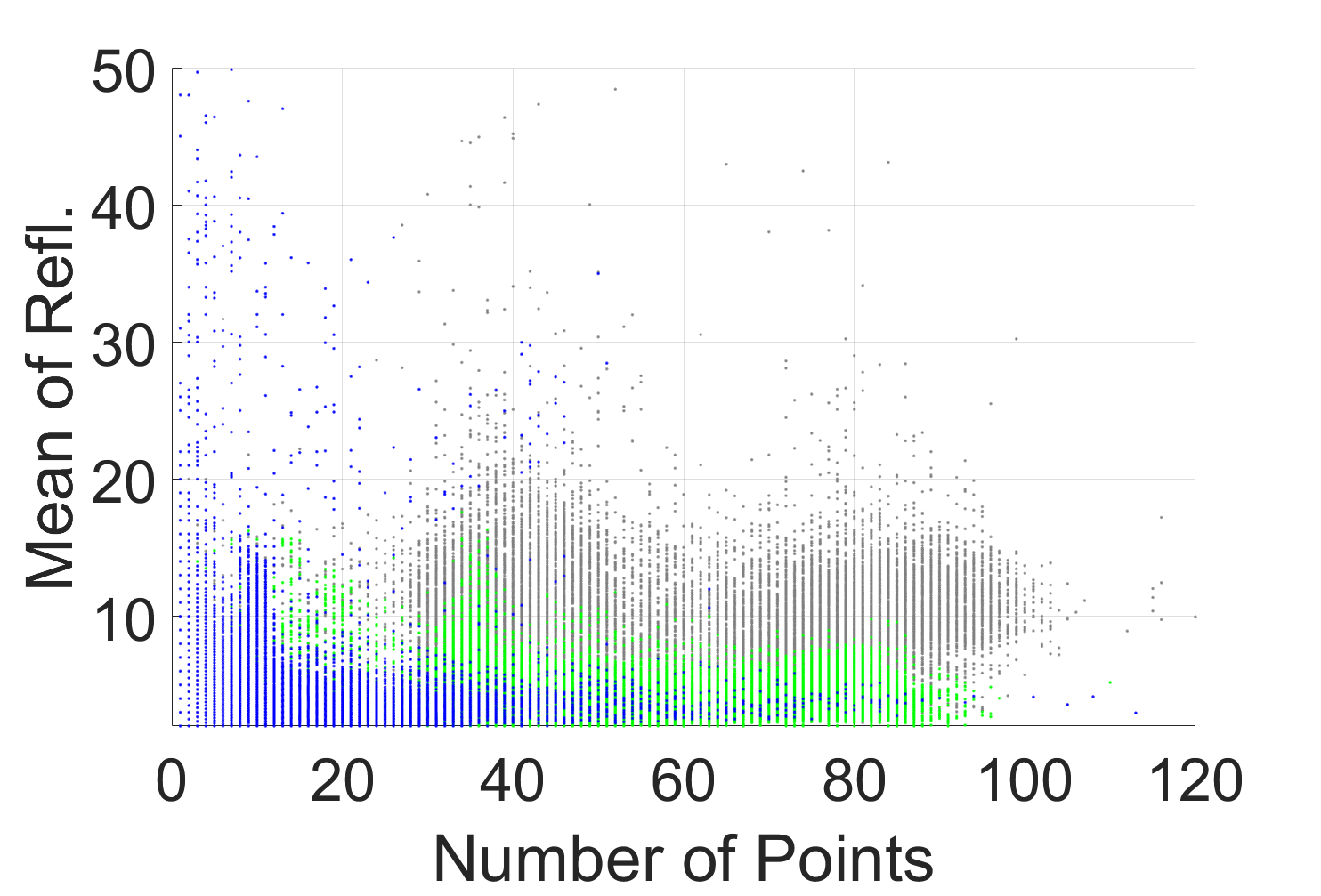}}
\subfloat[][RF Region Num.-Refl.]{\includegraphics[width=0.239\textwidth]{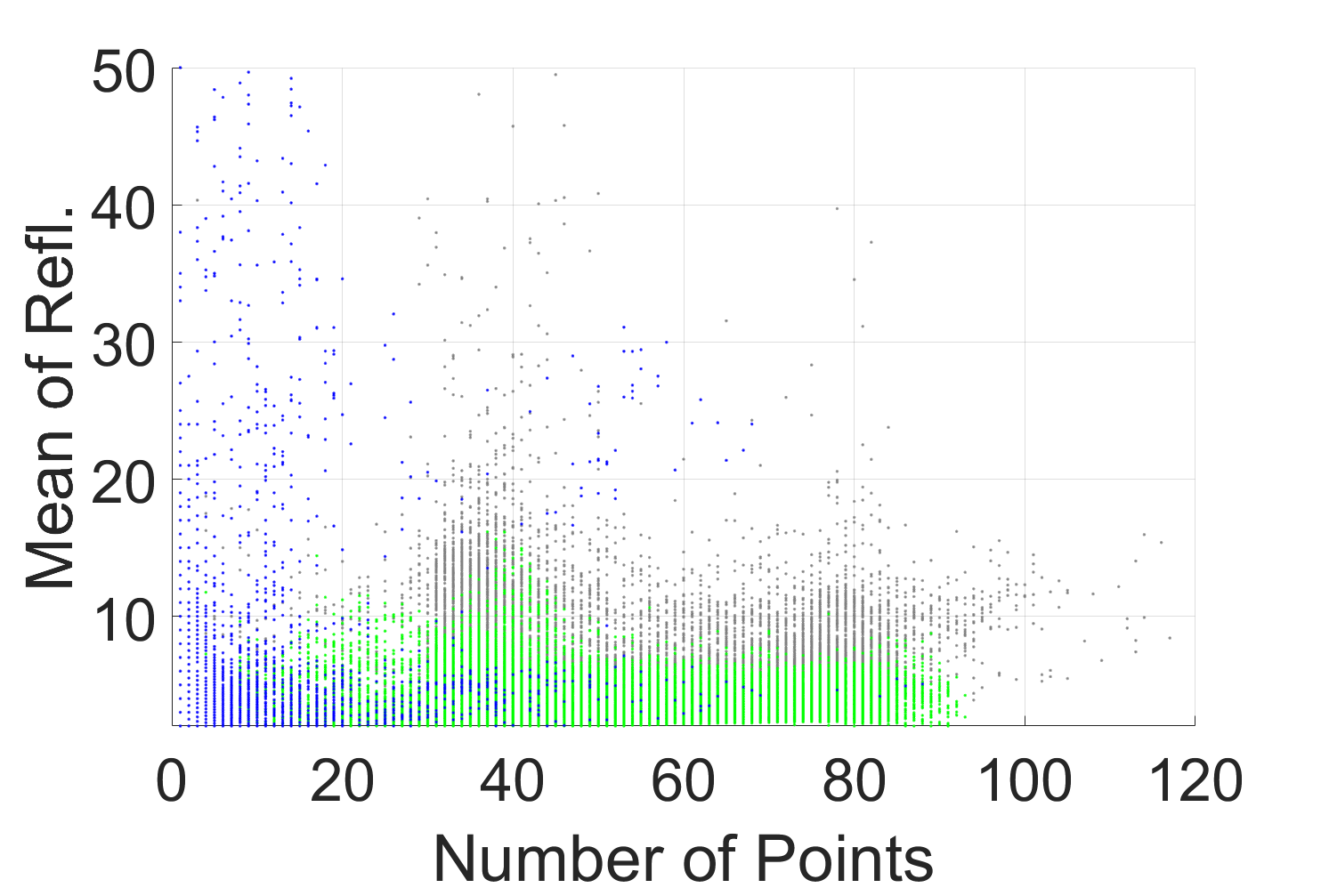}}
\subfloat[][LN Region Num.-Refl.]{\includegraphics[width=0.239\textwidth]{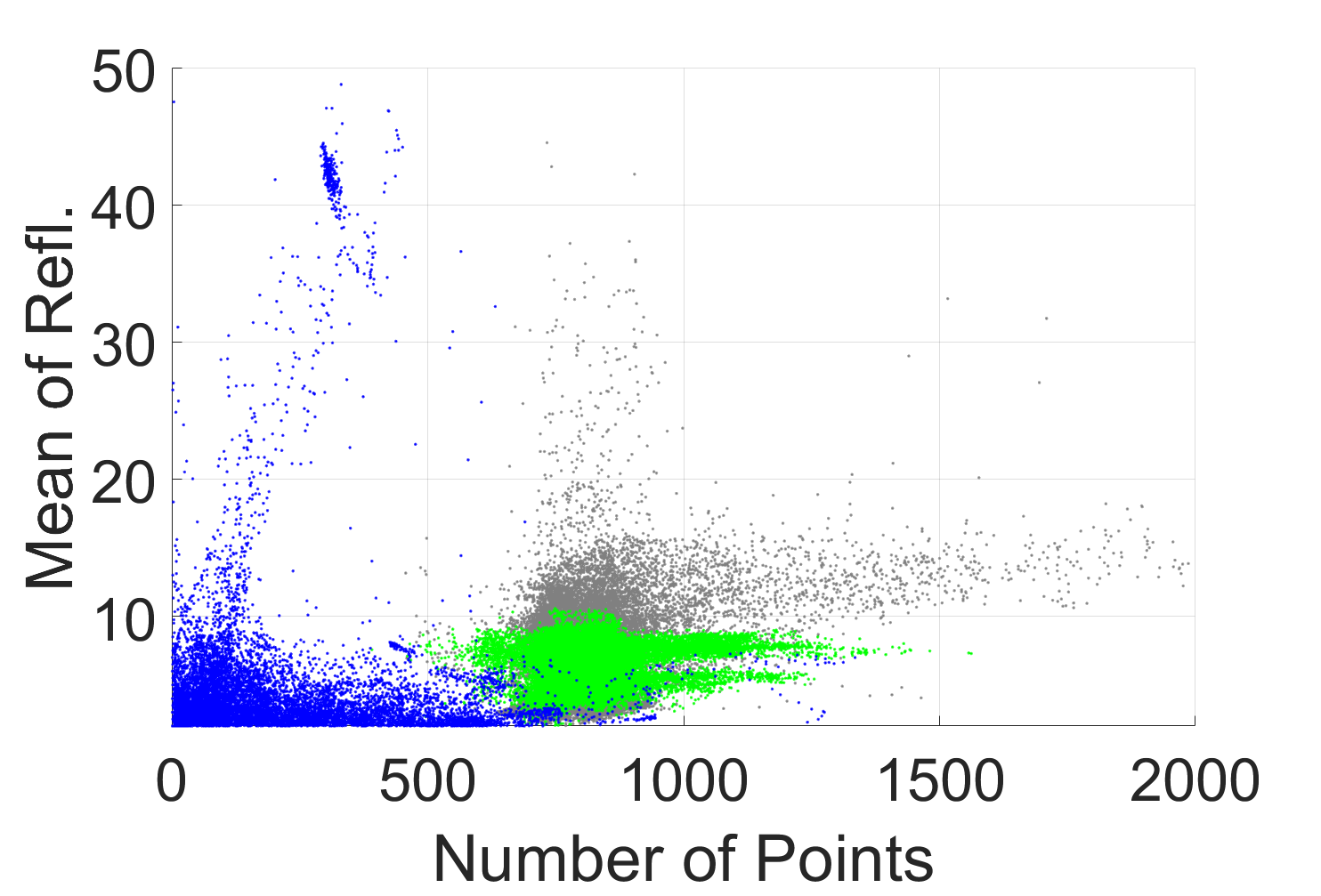}}
\subfloat[][RN Region Num.-Refl.]{\includegraphics[width=0.239\textwidth]{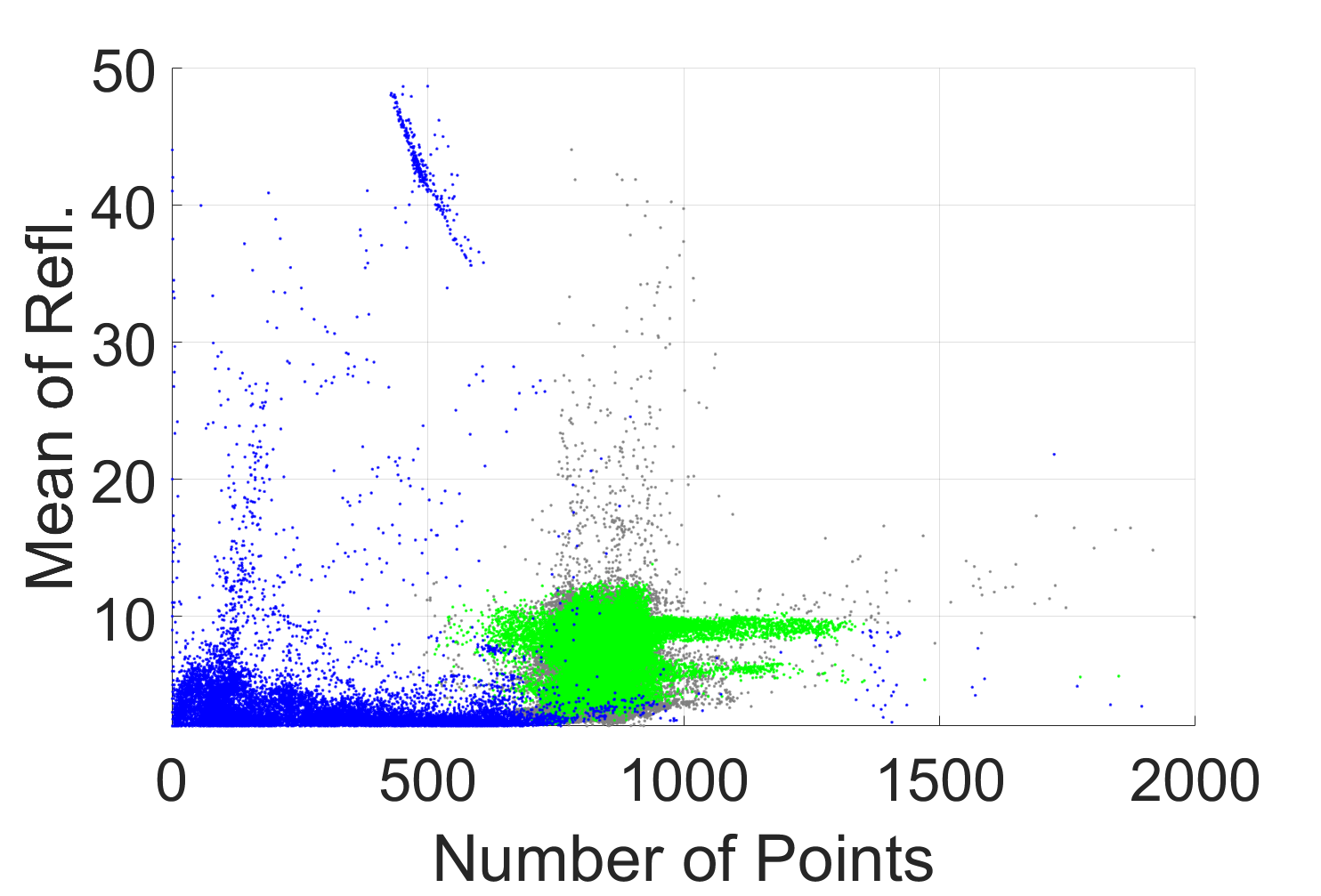}}
\hfil
\subfloat[][LF Region Time-Refl.]{\includegraphics[width=0.239\textwidth]{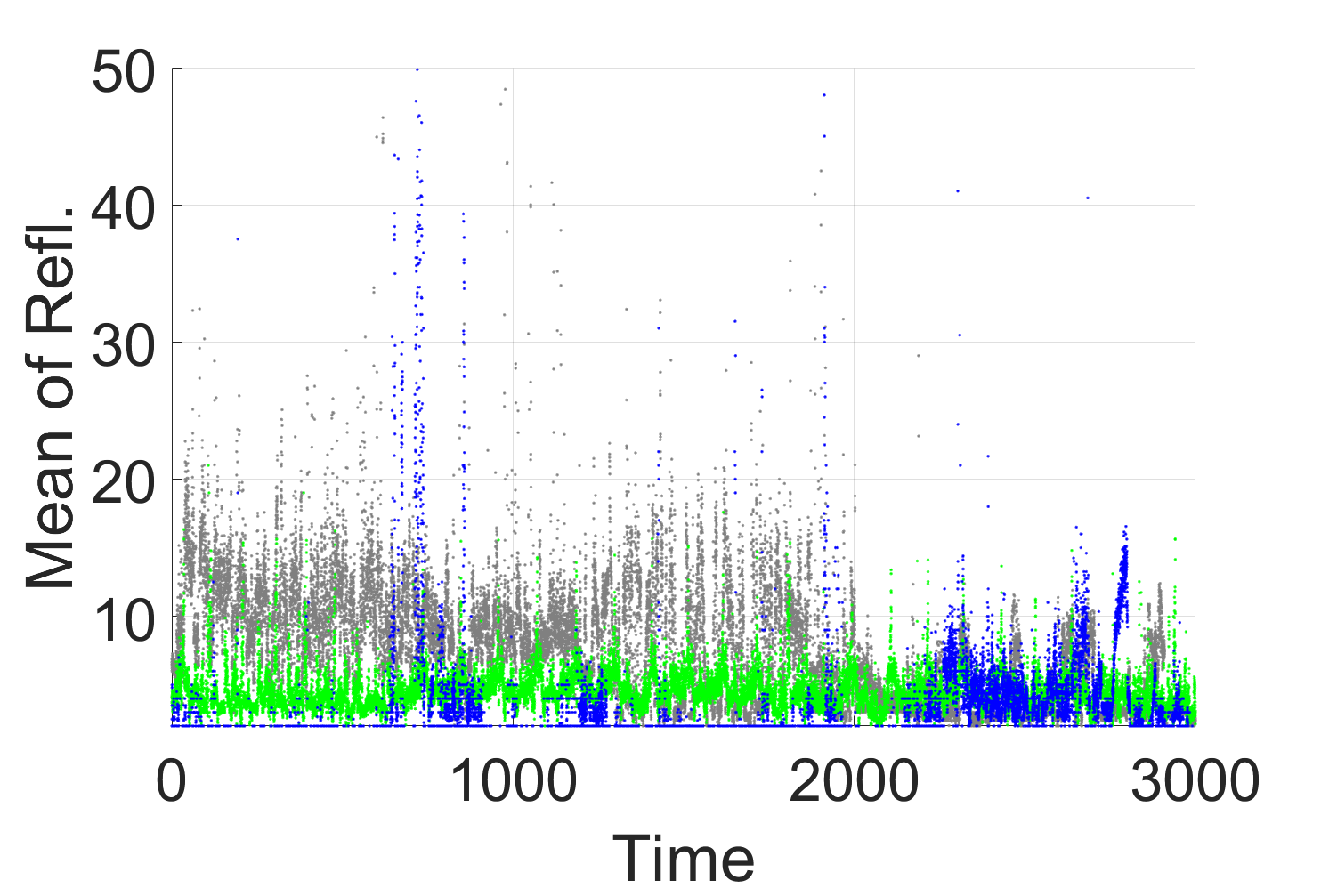}}
\subfloat[][RF Region Time-Refl.]{\includegraphics[width=0.239\textwidth]{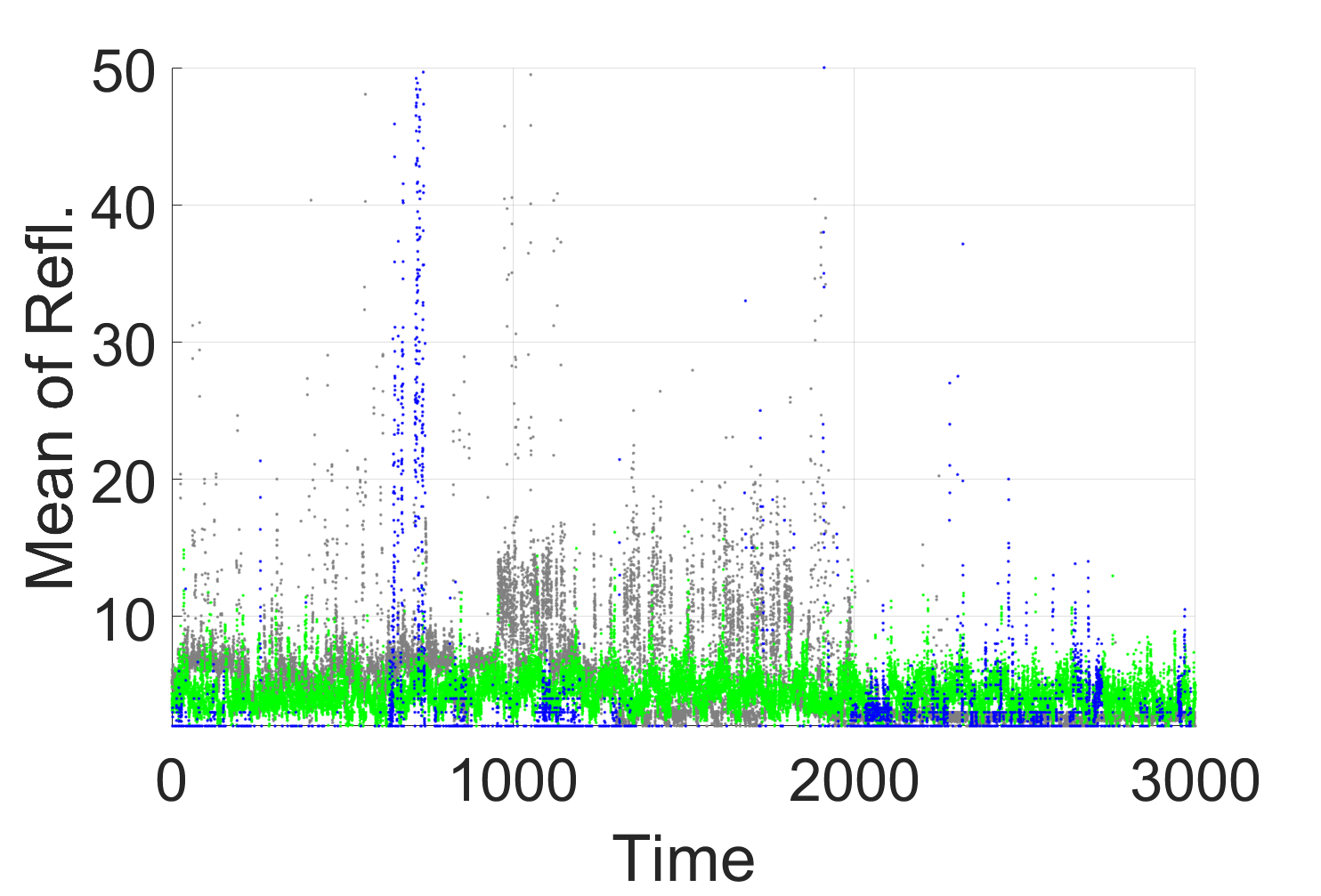}}
\subfloat[][LN Region Time-Refl.]{\includegraphics[width=0.239\textwidth]{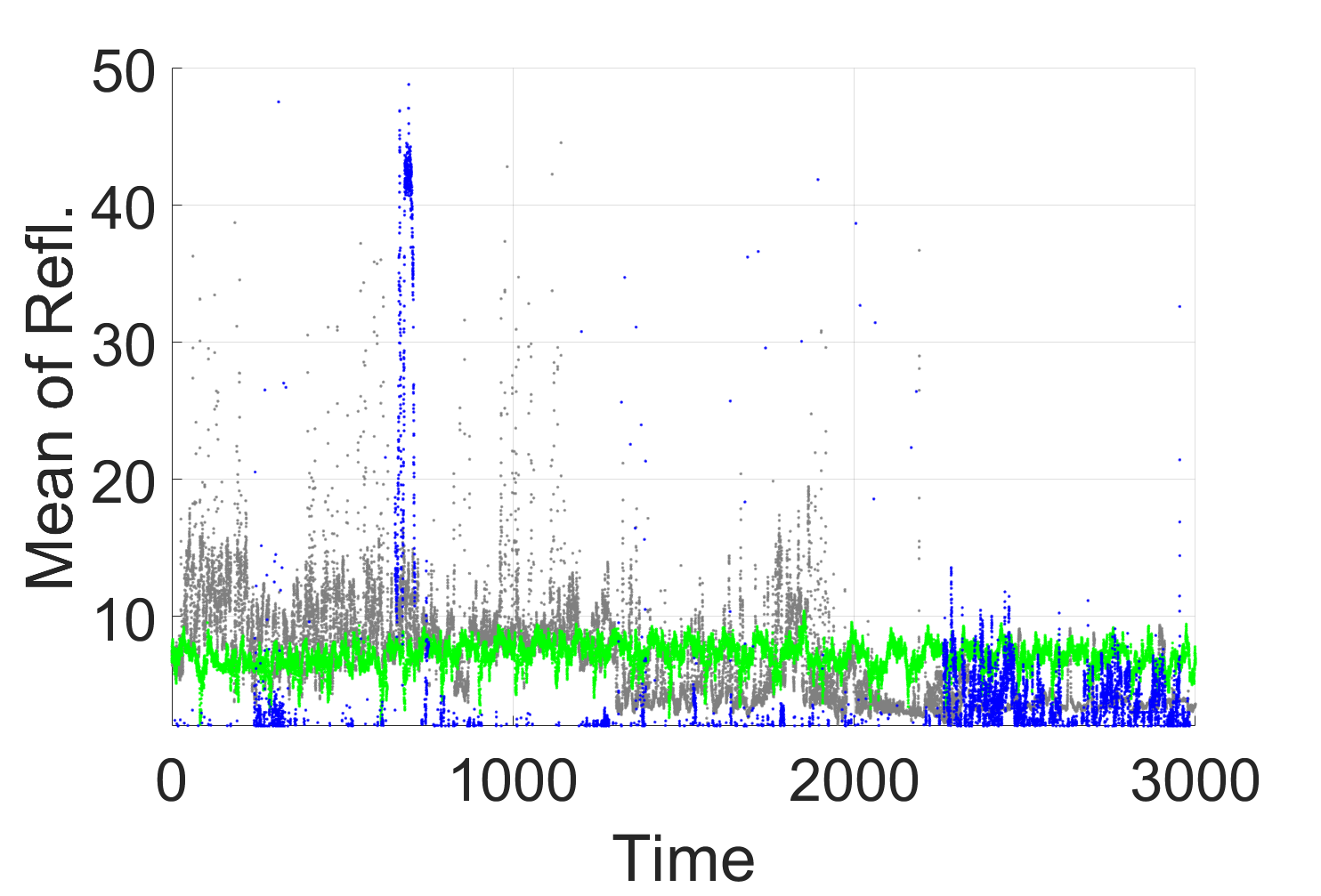}}
\subfloat[][RN Region Time-Refl.]{\includegraphics[width=0.239\textwidth]{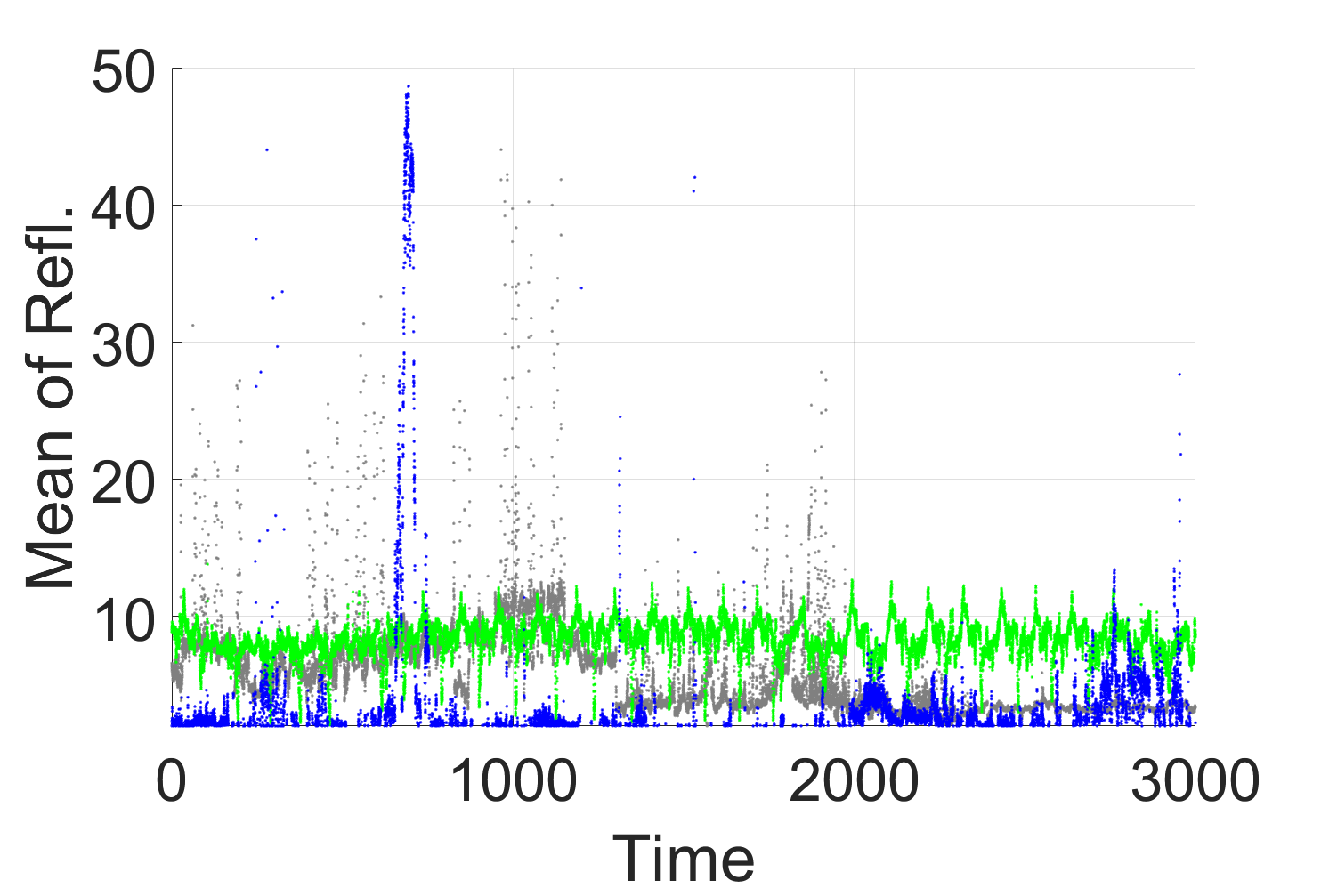}}
\caption{{Feature of the LiDAR data with each road condition. Gray indicated the dry condition, blue is the wet condition, and green is the snow condition. (a)-(d) are 3-dimension feature data (time vs number of points vs mean of reflecitivty), (e)-(h) are 2-dimension feature data (number of points vs mean of reflectivity), and (i)-(j) are 2-dimension feature data (time vs mean of reflectivity).}}
\label{fig:section_num_mean}
\end{figure*}
\begin{figure*}[t]
\centering
\subfloat[][Dry: Time-Num.-Refl.]{\includegraphics[width=0.239\textwidth]{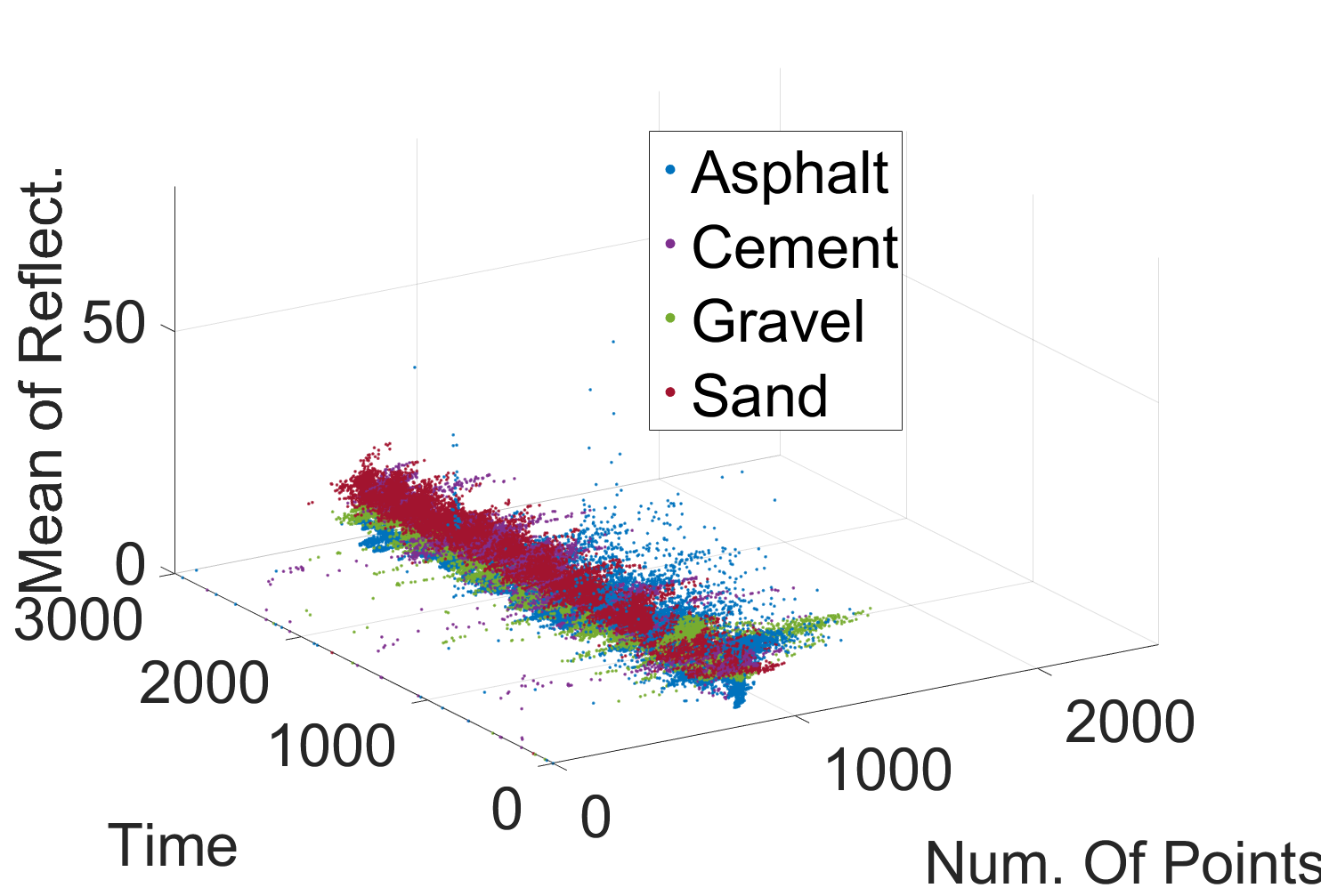}}
\subfloat[][Dry: Num.-Refl.]{\includegraphics[width=0.239\textwidth]{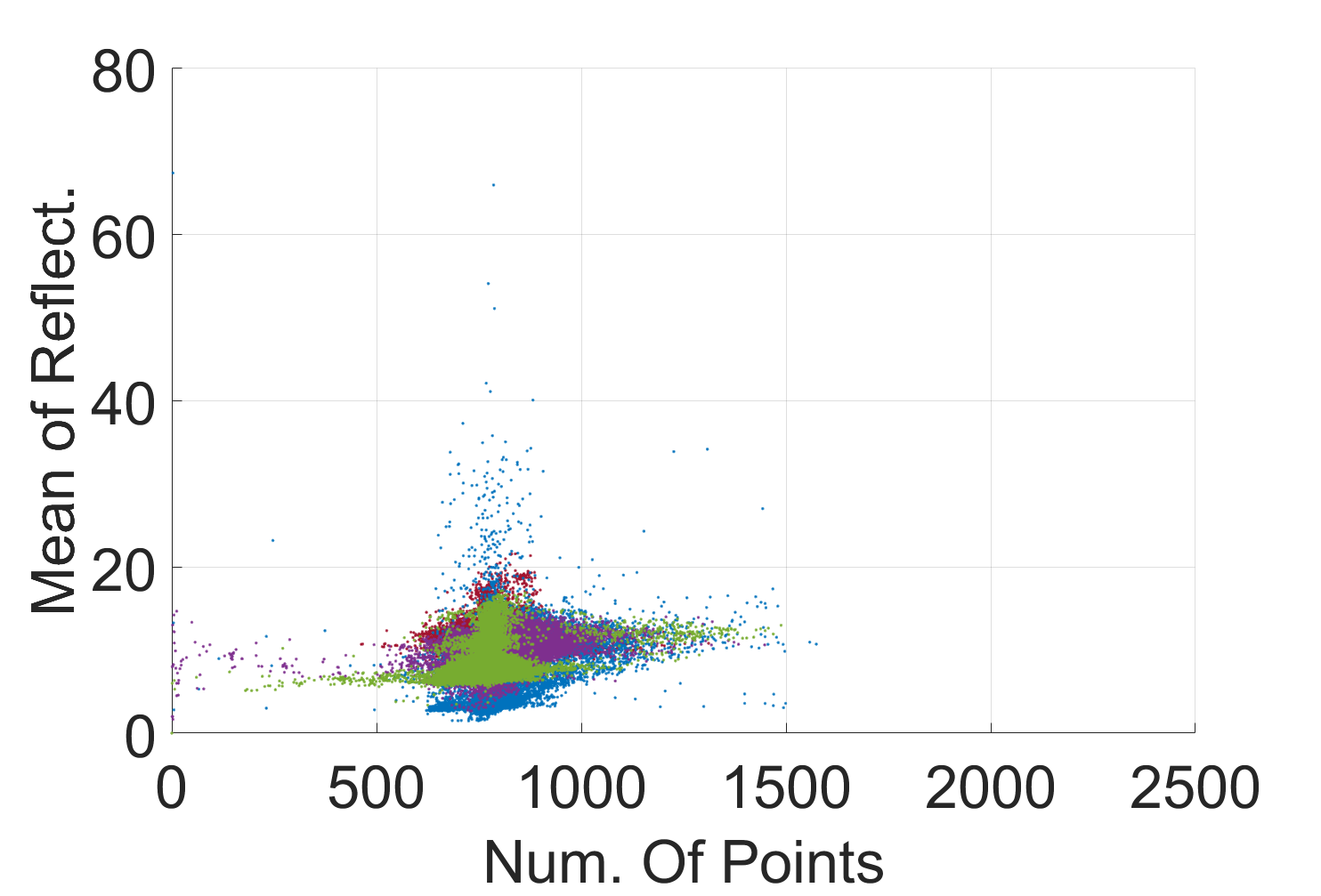}}
\subfloat[][Dry: Time-Refl.]{\includegraphics[width=0.239\textwidth]{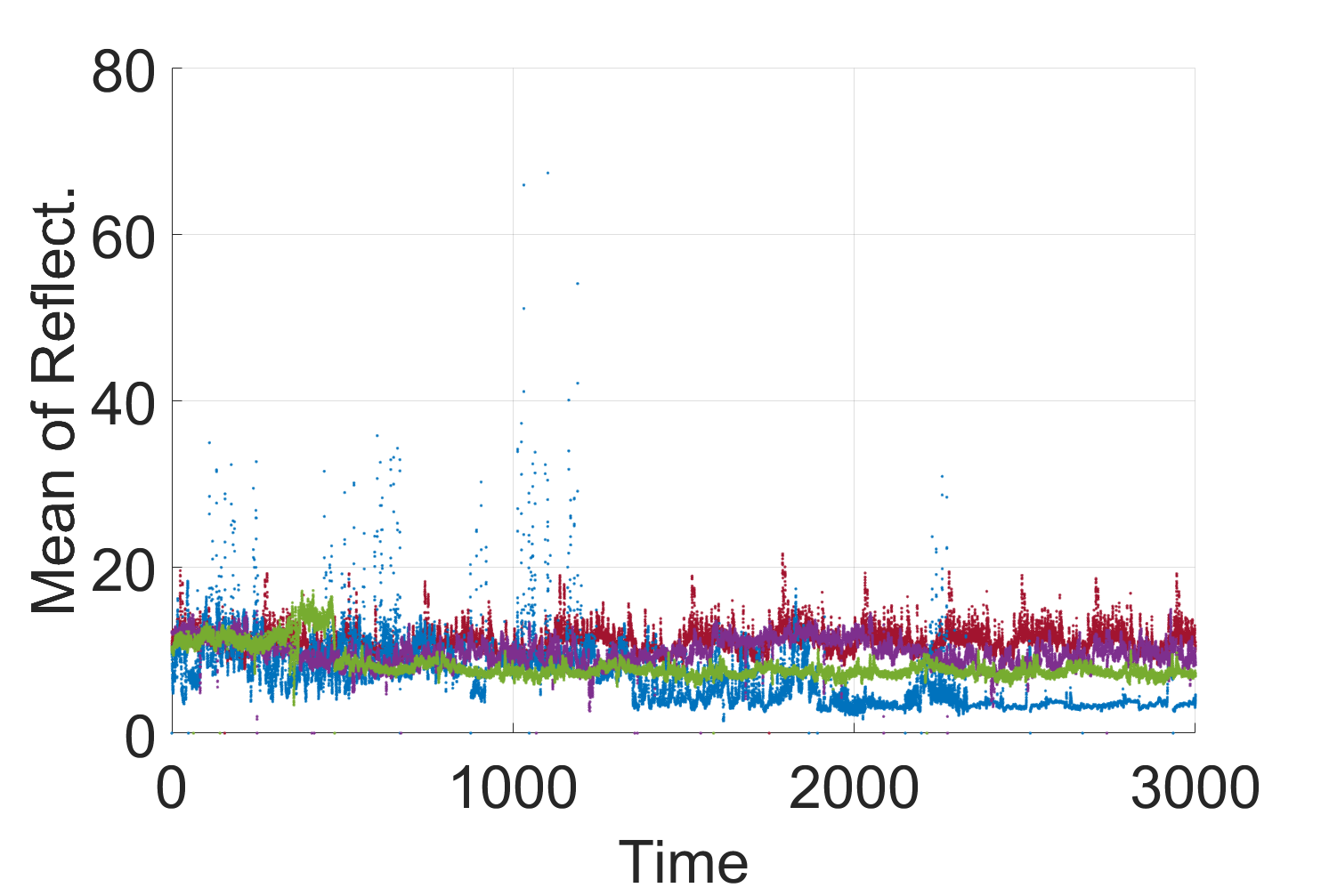}}
\subfloat[][Dry: Time-Num.]{\includegraphics[width=0.239\textwidth]{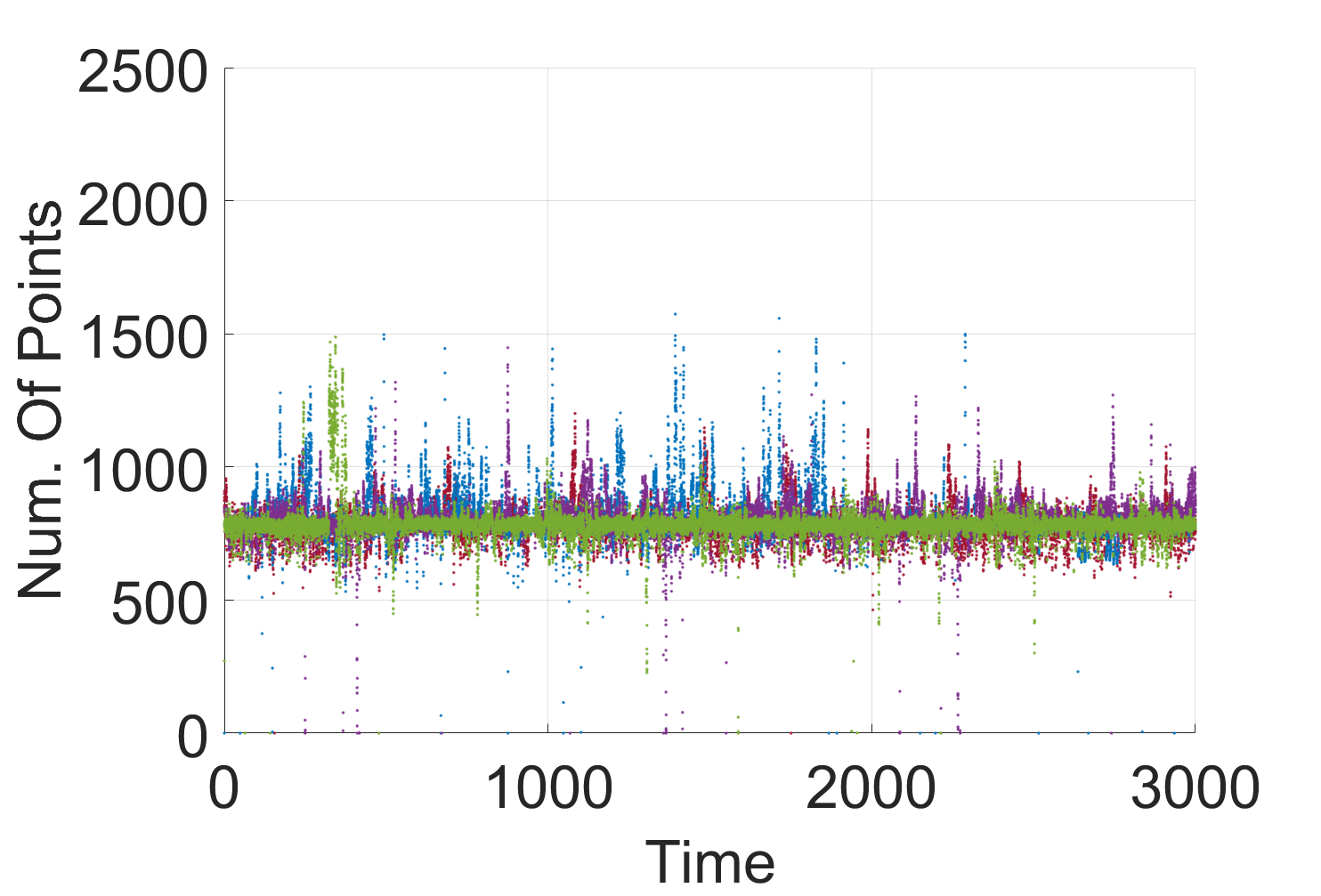}}
\hfil
\subfloat[][Wet: Time-Num.-Refl.]{\includegraphics[width=0.239\textwidth]{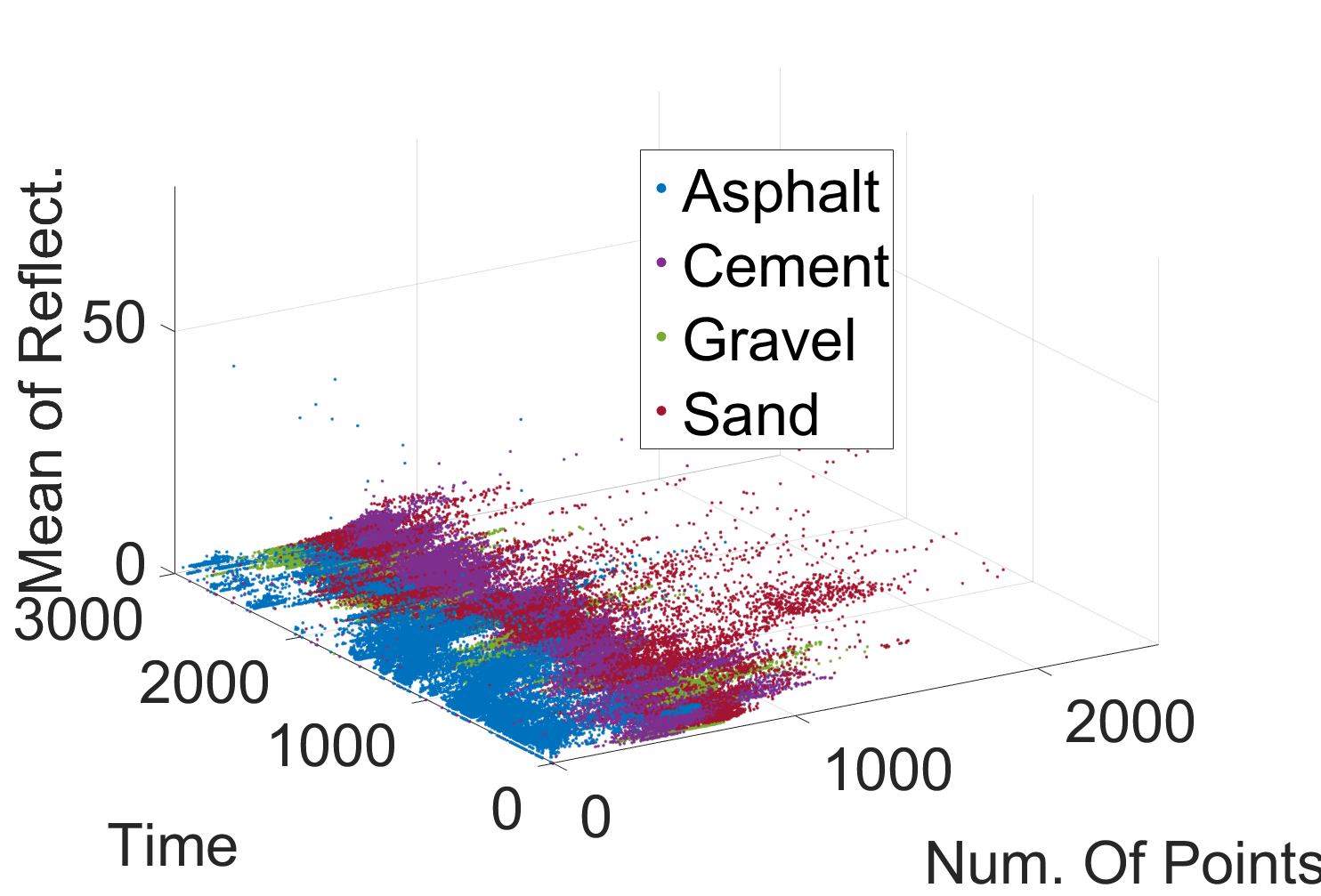}}
\subfloat[][Wet: Num.-Refl.]{\includegraphics[width=0.239\textwidth]{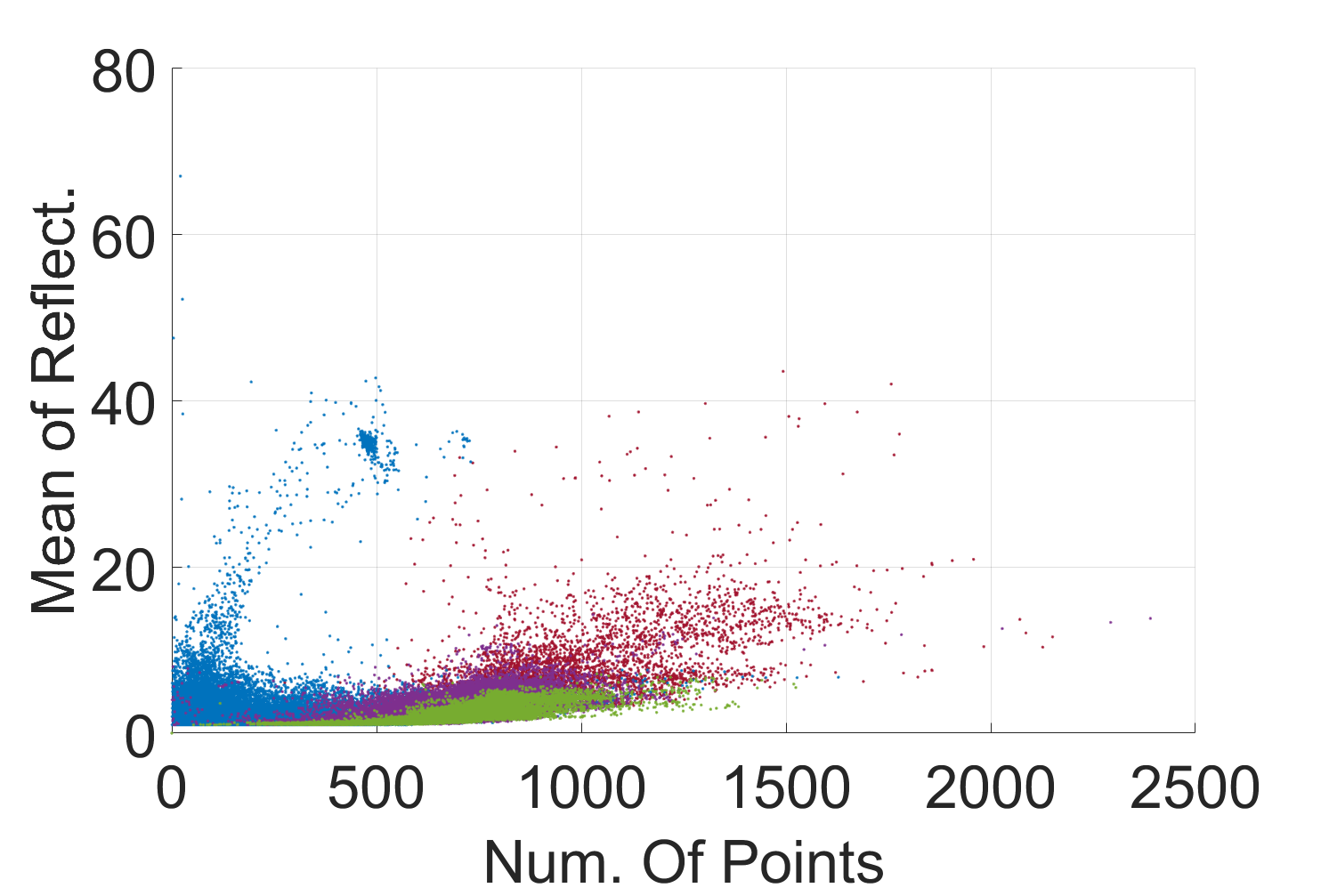}}
\subfloat[][Wet: Time-Refl.]{\includegraphics[width=0.239\textwidth]{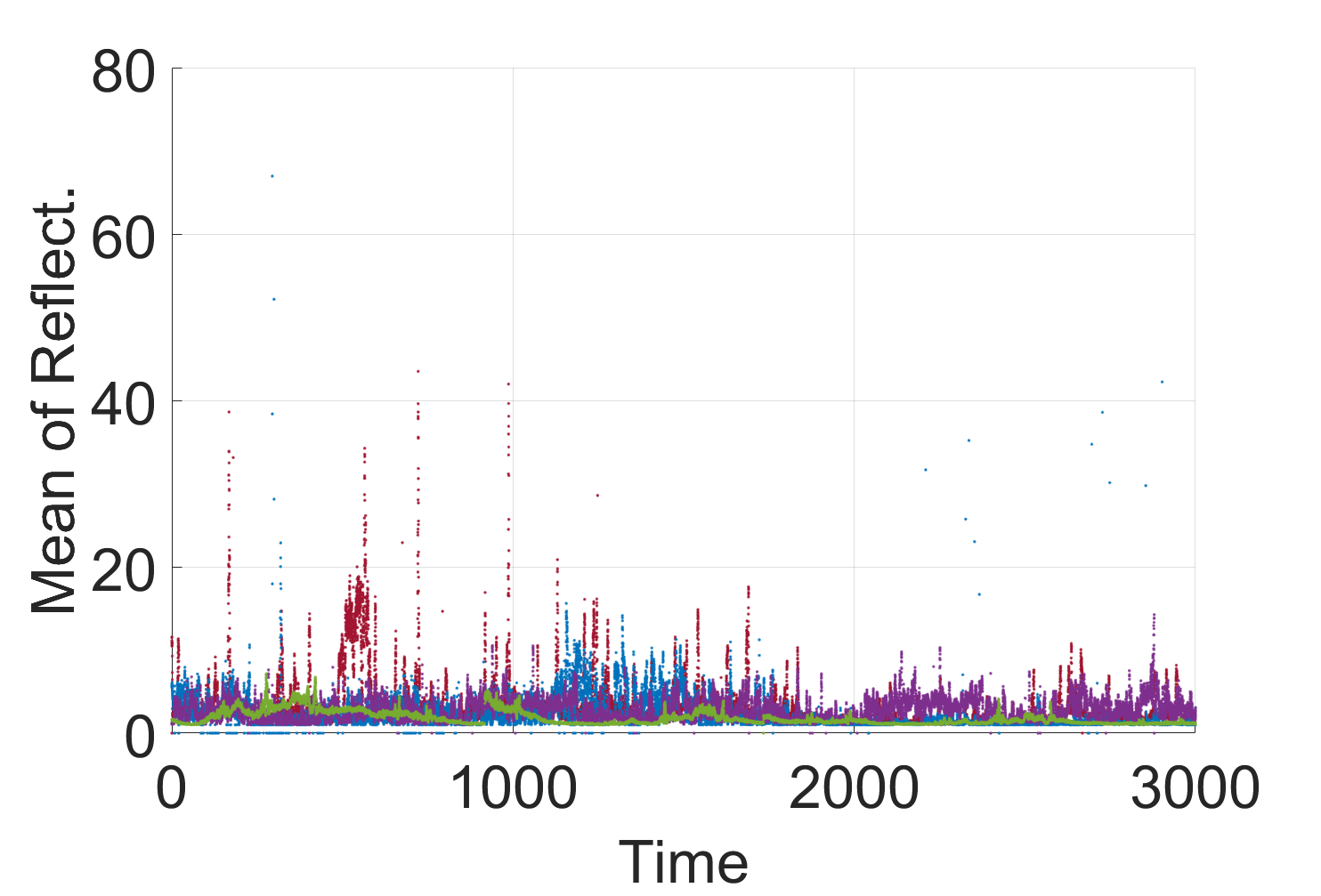}}
\subfloat[][Wet: Time-Num.]{\includegraphics[width=0.239\textwidth]{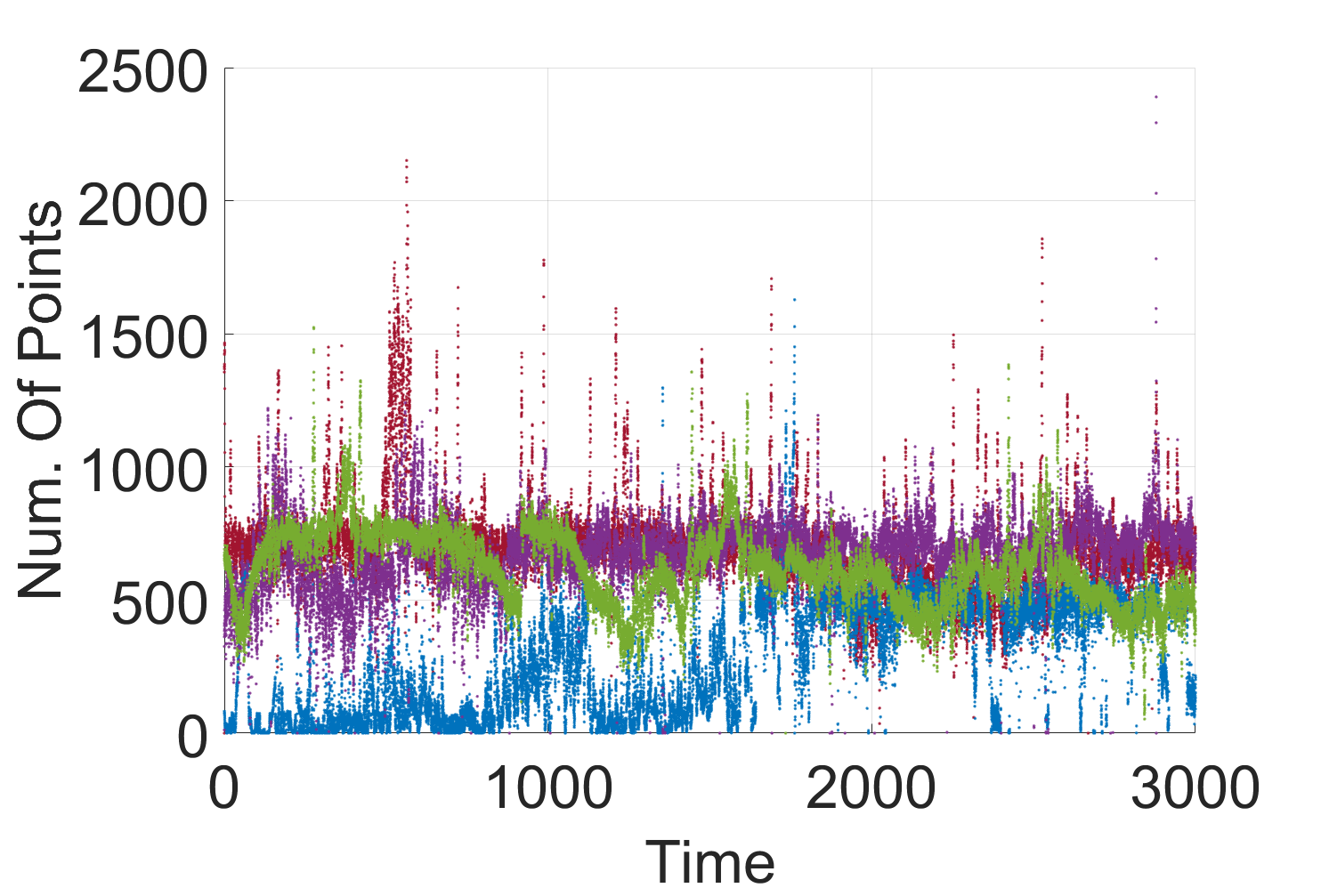}}
\caption{{Feature of the LiDAR data with each road type. (a)-(d) are the feature of the road type on dry road, and (e)-(h) are the feature of the road type on wet road.}}
\label{fig:type_feature}
\end{figure*}
Data processing is required to obtain the feature data. The first step is to select the feature vectors from the LiDAR sensor. One of the feature vectors is the mean value of the reflectivity information obtained during one rotation of the LiDAR in each region. Another feature vector is the total number of point clouds obtained during one rotation of the LiDAR in each region. The reason for exploiting this information is that the reflectivity and the number of point clouds returned to the receiver of LiDAR are different according to the roughness of the road surface~\cite{sebastian2021range}.
Regular reflection can occur when the road surface is polished smooth on a wet or snow road, resulting in few point clouds. However, mirror reflection occurs on the surface of the wet road, and point clouds return immediately so that some samples show high reflectivity. On the other hand, the number of point clouds and reflectivity may have been high on the dry road due to an irregular reflection~\cite{aki2016road}. To utilize this property of LiDAR data, the reflectivity and the number of point clouds are considered feature vectors.
The feature data of the LiDAR are illustrated in Fig.~\ref{fig:section_num_mean}. It shows that the distribution of features varies depending on the condition of the road.
Similar to the method of analyzing each road condition, the LiDAR data feature by type of road was analyzed.
Figure~\ref{fig:type_feature}~(a)-(d) show the feature of each type of road with the dry condition, and Figure~\ref{fig:type_feature}~(d)-(e) show the feature of each type of road with the wet condition. Since the roughness of the road surface is different depending on the type of road, it can be seen that reflectivity and the number of points are different. In order to analyze the area close to the vehicle mainly covered in this paper, the characteristic of the \emph{LN region} is represented.

The second step is to select the longitudinal speed obtained from the in-vehicle sensor as a feature vector. Most studies considered the characteristics of LiDAR, which varies depending on the road surface, but it is necessary to consider the feature that changes as the vehicle drives. Indeed, the data of LiDAR mounted on the vehicle is affected by vehicle speed since various motions occur as the vehicle moves~\cite{merriaux2017movingcar}. Therefore, the vehicle speed is selected as another feature vector to utilize the information of LiDAR in which the vehicle speed was considered together. The ablation study for selecting and not selecting the longitudinal speed as a feature vector will be discussed in Section~\ref{sec:ExperimentResult}.
Since the sample rate of the in-vehicle sensor is faster than that of LiDAR, unnecessary repetitive data for selecting the feature data must be prevented so that under-sampling of the in-vehicle sensor data is conducted with a sample rate of LiDAR (100 ms).

Finally, the feature vectors are stacked from the current time to the past one second using the time-windowing method, which causes the feature vector to contain the temporal information. Then an input vector of DNN is generated by this feature vector and entered into the neural network as shown in Fig.~\ref{fig:overall}. To create the target data, we utilized the one-hot-encoding method to classify the nine classes, for example, $\begin{bmatrix} 1 & 0 & 0 & \hdots & 0\end{bmatrix}^T \in \mathbb{R}^{n_{c}\times1}$ denotes dry asphalt road, and $\begin{bmatrix} 0 & 1 & 0 & \hdots & 0\end{bmatrix}^T \in \mathbb{R}^{n_{c}\times1}$ denotes dry cement road, where $n_{c}$ is the number of classes.
\section{Deep Neural Network with Spatiotemporal Information}
\subsection{Deep Neural Network and Optimization}
\begin{figure*}[t]
\centering
{\includegraphics[width=0.9\textwidth]{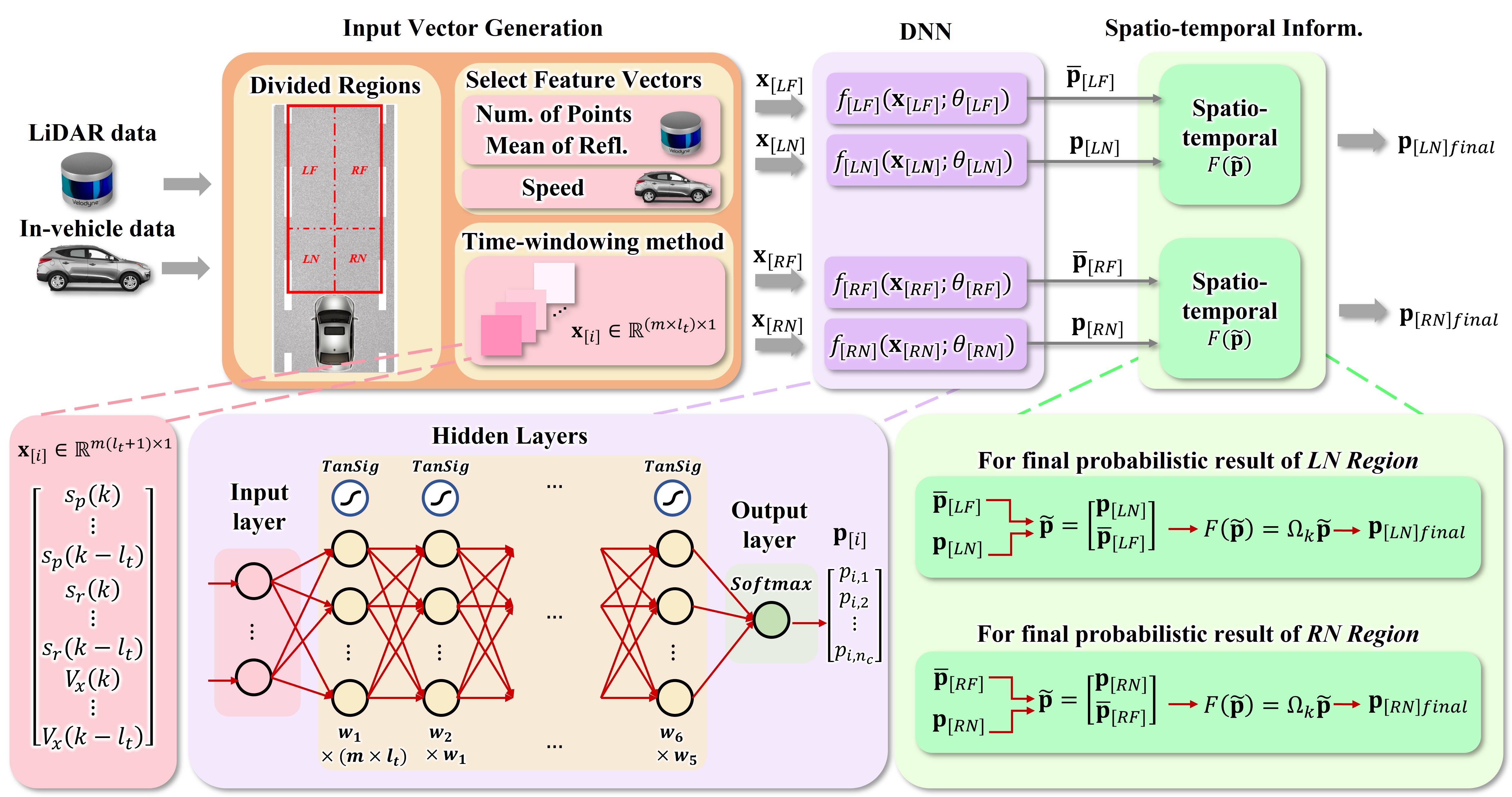}}\caption{
{Overall structure of the proposed algorithm. The final result $\textbf{p}_{[LN]final}$ and $\textbf{p}_{[RN]final}$ is updated through the function $F(\tilde{\textbf{p}})$ by using the concatenated vector $\tilde{\textbf{p}}$ that stack the current result and past results over $k-l_{s}$ to $k$ step. ($m=3,\,\,l_t=10,\,\,l_s=5,\,\,n_c=9,\,\,i\in\{LN,\,RN,\,LF,\,RF\}$)}}
\label{fig:overall}
\end{figure*}
\begin{figure*}[t]
\centering
{\includegraphics[width=0.9\textwidth]{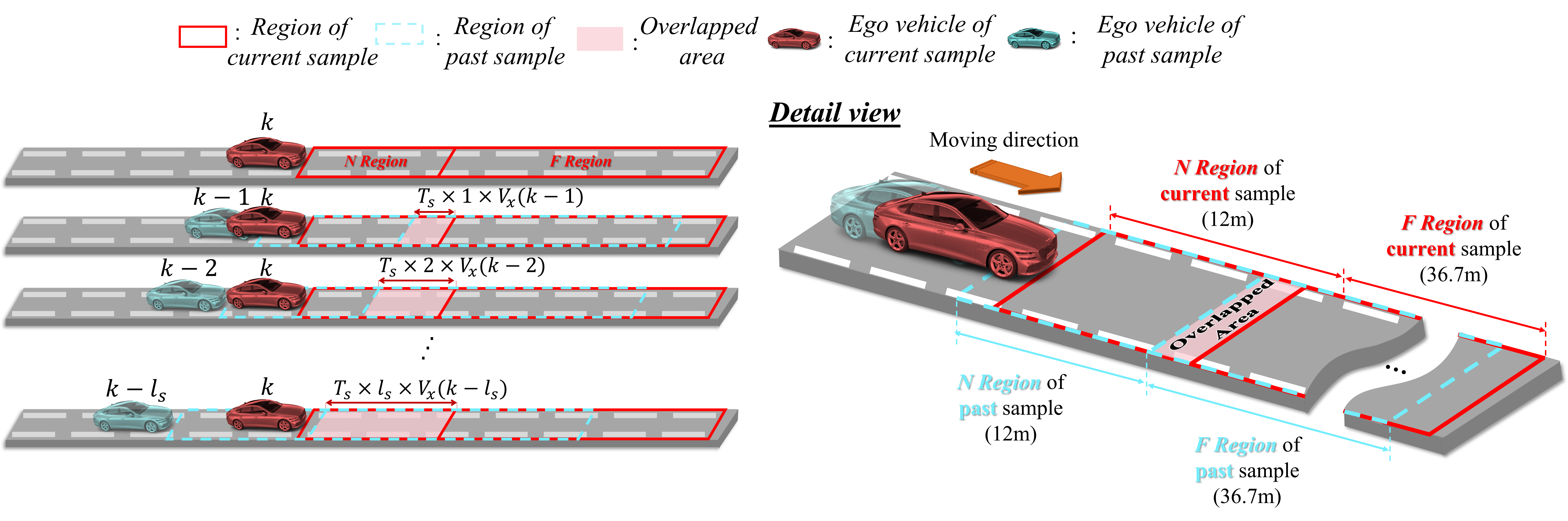}}
\caption{{As the vehicle moves forward at the speed $V_x$, the yellow hatched region of the current sample time, and the blue hatched region of the past $k-1$ sample time overlap by a distance of $T_s\times1\times V_x(k-1)$ (Blue region). In the same way, the ratio of overlapping in the $k-2,\,k-3,\,...,\,k-l_s$ sample time can be determined. The road condition of the yellow hatched region can be classified by using the spatiotemporal information with the past results of the blue hatched region. In this paper, we set the parameter as $l_s=5$, $T_s=0.1s$.}}
\label{fig:overlap}
\end{figure*}
We utilize the deep neural network to classify the road surface. The overall configuration of the network is shown in Fig.~\ref{fig:overall}. The network consists of six hidden layers and one output layer. The transfer function for each layer is set to a non-linear function, e.g., the tangent-sigmoid function, since it has a very smooth curvature and does not cause gradient exploration affected by a sharp change in slope. The number of neurons in each layer is set to $100$, $80$, $40$, $40$, $20$, and $10$, respectively. The activation function of the output layer is utilized using the softmax method, in which the output is obtained as the probability corresponding to each class. Here, we design the neural network for each region, meaning there are four neural networks for classifying the road conditions at each region.

The scaled conjugate gradient (SCG) backpropagation algorithm was selected to optimize the network. SCG is a Conjugate Gradient Algorithm (CGA) variation ~\cite{moller1993scaled}. Since most CGAs conduct a line search for every iteration, they require complex computation and a long calculation time. SCG with a step-size scaling mechanism was developed to overcome this problem. The basic idea of the SCG is to combine the model-trust region approach used in the Levenberg-Marquardt algorithm with the CGA. Thus, the SCG was used to optimize classification problems~\cite{beale1992neural}. In DNN, the optimization is conducted by minimizing the error between the target data and the neural network's output. Let the error function in the training set can be written as
\begin{gather}
\begin{split}
&E_{[i]}=f_{[i]}(\textbf{x}_{[i]} ; \mathbf{\theta}_{[i]})-T_{[i]},\\ &i\in\{LF,RF,LN,RN\}
\end{split}
\end{gather}
where $i$ is the index of the networks for each region, $f_{[i]}(\textbf{x}_{[i]};\theta_{[i]})$ is the output of the neural network, $T_{[i]}$, $\textbf{x}_{[i]}$, and $\theta_{[i]}$ are the target data, the selected feature vector inputs, and weight parameters for the $i\in\{LF,RF,LN,RN\}$, respectively.
The loss function used for training the network is the mean square error (MSE) function, as follows:
\begin{gather}
\label{eq:lf}
L_{[i]}^n=\frac{1}{N_{b}}\sum_{l=(n-1)N_{b}+1}^{n\cdot N_{b}}{E_{[i]}^l}^T\cdot E_{[i]}^l+\frac{\lambda}{2}||\theta_{[i]}||_{2}^{2}
\end{gather}
where $n$ and $N_{b}$ are the index and size of the mini-batch, respectively. Here, we use the regularization strategy to prevent overfitting that adds the $L^{2}$ parameter norm penalty, with hyper-parameter $\lambda$,  $\frac{\lambda}{2}||\theta_{[i]}||_{2}^{2}$ to MSE function~\cite{fellow2016deep}. Now, the SCG algorithm is utilized, and the optimal parameter of the network is updated as follows:
\begin{gather}
\theta_{[i]}^n=\theta_{[i]}^{n-1}+\gamma_{[i]}^n\nabla_{[i]}^n L_{[i]}^n
\end{gather}
where $\gamma_{[i]}^n$ is a step size, and $\nabla_{[i]}^n$ is a conjugate direction, respectively. The step size is calculated by the algorithm from~\cite{moller1993scaled}. The final optimal parameter can be obtained through the optimization process as
\begin{gather}
\theta_{[i]}^*=\arg\min_{\theta_{[i]}^{n}}(L_{[i]}^{n}(f_{[i]}(\textbf{x}_{[i]} ; \mathbf{\theta}_{[i]}),T_{[i]})).
\end{gather}

\subsection{Spatiotemporal Formulation}
To generate the input vector, we define the concatenated vectors from $k-l_{t}$ to $k$ steps $\bar{\textbf{s}}_{p}(k|k-l_{t})$, $\bar{\textbf{s}}_{r}(k|k-l_{t})$, and $\bar{\textbf{V}}_{x}(k|k-l_{t})$ such as
\begin{gather}
\begin{split}
&\bar{\textbf{s}}_{p}(k|k-l_{t})=[s_{p}(k),\,s_{p}(k-1),\,\hdots\,,s_{p}(k-l_{t})]^{T}\\
&\bar{\textbf{s}}_{r}(k|k-l_{t})=[s_{r}(k),\,s_{r}(k-1),\,\hdots\,,s_{r}(k-l_{t})]^{T}\\
&\bar{\textbf{V}}_{x}(k|k-l_{t})=[V_{x}(k),\,V_{x}(k-1),\,\hdots\,,V_{x}(k-l_{t})]^{T}
\end{split}
\end{gather}
where $l_{t}$ is the stacking step for the time windowing method, $s_{p}$ is the number of points, $s_{r}$ is the mean of the reflectivity, and $V_{x}$ is the longitudinal speed of ego-vehicle, respectively.
Let $\textbf{x}_{[i]}\in\mathbb{R}^{m(l_{t}+1)\times 1}$ be the input vector for $i\in\{LF, RF, LN, RN\}$ where $m$ is the number of feature.
Then, we can present the input vector of the network as
\begin{gather}
\textbf{x}_{[i]}=
\begin{bmatrix}
\bar{\textbf{s}}_{p}(k|k-l_{t})\\
\bar{\textbf{s}}_{r}(k|k-l_{t})\\
\bar{\textbf{V}}_{x}(k|k-l_{t})
\end{bmatrix}\in\mathbb{R}^{m(l_{t}+1)\times 1}.
\end{gather}

We want to estimate the probability that $P(y=N|\textbf{x}_{[i]})$ for each value of $N=1,\hdots,n_{c}$ to utilize the spatiotemporal formulation where $n_{c}$ is the number of classes. In other words, we want to obtain the class label's probability distribution output, taking on each $N$ different possible value. To this end, we added the {\it{softmax}} function, which normalizes the probability distribution to the exponentials, to the output layer of the network model~\cite{fellow2016deep}. Using this exponential property, the probability can be presented as
\begin{gather}
\begin{split}
P(y=N|\textbf{x}_{[i]})=\frac{\exp(\textbf{x}_{[i]}^{T}\theta_{[i]}^{(N)})}{\sum_{l=1}^{n_{c}}\exp(\textbf{x}_{[i]}^{T}\theta^{(l)}_{[i]})}
\end{split}
\end{gather}
where $\theta_{[i]}^{(\cdot)}$ are the weighting vectors of the optimized network.
Therefore, the network output has a $n_c$-dimensional vector, whose elements sum to 1, giving us $n_{c}$ predicted probabilities. Let $\textbf{p}_{[i]}\in\mathbb{R}^{n_{c}\times 1}$ be the probability distribution output from the optimized neural network for $i\in\{LF, RF, LN, RN\}$ through the {\it{softmax}} layer, then it can be presented as
\begin{gather}\label{eq:softmax_result}
\begin{split}
\textbf{p}_{[i]}&=
\begin{bmatrix}
P(y=1|\textbf{x}_{[i]};\theta_{[i]})\\
P(y=2|\textbf{x}_{[i]};\theta_{[i]})\\
\vdots\\
P(y=n_{c}|\textbf{x}_{[i]};\theta_{[i]})
\end{bmatrix}\\
&=\frac{1}{\sum_{l=1}^{n_{c}}\exp(\textbf{x}_{[i]}^{T}\theta_{[i]}^{(l)})}
\begin{bmatrix}
\exp(\textbf{x}_{[i]}^{T}\theta_{[i]}^{(1)})\\
\exp(\textbf{x}_{[i]}^{T}\theta_{[i]}^{(2)})\\
\vdots\\
\exp(\textbf{x}_{[i]}^{T}\theta_{[i]}^{(n_{c})})
\end{bmatrix}.
\end{split}
\end{gather}
Here, we introduce concatenated vector $\tilde{\textbf{p}}(k)\in\mathbb{R}^{n_c(l_{s}+1)\times1}$ with stacking step $l_{s}$ by stacking the current probabilistic result of \{\emph{LN Region, RN Region}\} and past probabilistic result of \{\emph{LF Region, RF Region}\} over $k-l_{s}$ to $k$ step as
\begin{gather}\label{eq.stack_output}
\begin{split}
\tilde{\textbf{p}}(k)=
\begin{bmatrix}
\textbf{p}_{[i]}(k) \\
\bar{\textbf{p}}_{[j]}(k-1|k-l_{s})
\end{bmatrix}
\end{split}
\end{gather}
where
\begin{gather*}
\bar{\textbf{p}}_{[j]}(k-1|k-l_{s})=
\begin{bmatrix}
\textbf{p}_{[j]}(k-1)\\
\textbf{p}_{[j]}(k-2)\\
\vdots \\
\textbf{p}_{[j]}(k-l_{s})
\end{bmatrix}\in\mathbb{R}^{n_{c}l_{s}\times 1},\\
i\in\{\emph{LN, RN}\},~j\in\{\emph{LF,~RF}\}.
\end{gather*}
Here, the spatiotemporal function with concatenated vector $F(\tilde{\textbf{p}}(k))$ is defined as follows:
\begin{gather}
\begin{split}
F(\tilde{\textbf{p}}(k)):=(\Omega_{k} \otimes I_{nc})\tilde{\textbf{p}}(k)\in\mathbb{R}^{n_{c}\times 1}
\end{split}
\end{gather}
where $\otimes$ means the Kronecker product, $I_{n_c}\in\mathbb{R}^{n_{c} \times n_{c}}$ denotes $n_{c}\times n_{c}$ identity matrix and
\begin{gather*}
\Omega_{k}=
\begin{bmatrix}
\alpha_{k}& \alpha_{k-1}& \hdots & \alpha_{k-l_{s}}
\end{bmatrix}\in\mathbb{R}^{1\times (l_{s}+1)},
\end{gather*}
with the time-varying weight constants $\alpha_{l}$ for $l\in\{k,k-1,\hdots, k-l_{s}\}$ which are $\sum_{l=k}^{k-l_{s}}\alpha_{l}=1$ corresponding to each probabilistic result. Network output results of past \emph{LF region} and \emph{RF region} are used to obtain road conditions and types of \emph{LN region} and \emph{RN region} in the current step.

In order to utilize the spatiotemporal information, it is necessary to set weight constants, $\alpha_{k}$ on the DNN output probabilities. To this end, we utilize the length information of the overlapped area, as shown in Fig.~\ref{fig:overlap}.
Let the current time's vehicle speed be $V_x(k)$~m/s, and $T_s$ is the sample rate of the LiDAR sensor (100 ms).
When the vehicle moves forward at the longitudinal speed, $V_x$, the \emph{LF Region} in step $k-1$ overlaps with the \emph{LN Region} in step $k$ as the $T_s \times 1 \times V_x(k-1)$ distance. In general formulation, the \emph{LF Region} in step $k - l_s$ overlaps with the \emph{LN Region} in step $k$ as the $T_s \times l_s \times V_x(k - l_s)$ distance. Likewise, the \emph{RF Region} and \emph{RN Region} overlap in the same method.
By utilizing this spatial information, it is possible to determine the road condition and type in the \emph{LN Region} fused with the past probabilistic results of the \emph{LF Region} and in the \emph{RN Region} with the past probabilistic results of the \emph{RF Region}.
In~\eqref{eq.weight}, the weight matrix $\Omega_{k}(\cdot)$ dependent on $V_{x}$ can be determined as follows:
\begin{gather}\label{eq.weight}
\begin{split}
&\Omega_{k}(\bar{\textbf{V}}_{x}(k-1|k-l_{s}))\\
&\,\,\,=\frac{1}{\beta_{1}+\sum_{l=2}^{l_{s}}(l\cdot T_{s}\cdot V_{x}(k-l))}
\begin{bmatrix}
\beta_{1}\\
T_{s}V_{x}(k-1)\\
2T_{s}V_{x}(k-2)\\
\vdots\\
l_{s}T_{s}V_{x}(k-l_{s})
\end{bmatrix}^{T}
\end{split}
\end{gather}
Here, $\beta_1$
is the length of the \emph{LN Region}(or \emph{RN Region}), e.g., $12$, and the term $1/(\beta_{1}+\sum_{l=2}^{l_{s}}(l\cdot T_{s}\cdot V_{x}(k-l)))$ needs for normalization. Using this weight matrix, we can propose a spatial fusion method as follows:
\begin{gather}\label{eq:geometric}
\begin{split}
&\textbf{p}_{[i]final}(k)= F(\tilde{\textbf{p}}(k))\\
&i\in\{LN,RN\}.
\end{split}
\end{gather}
The overall process of the proposed method can be presented as Algorithm~\ref{ag.algorithm}.
\begin{algorithm}
\caption{Fusion algorithm}
\label{ag.algorithm}
\begin{algorithmic}[1]
\REQUIRE{Network model $f_{[i]}$, Network input $\textbf{x}_{[i]}$, Longitudinal speed $V_{x}$, Sample time $T_{s}$, Number of validation dataset $N_{val}$}, Number of stacking step $l_{s}$
\REQUIRE{$i\in\{\emph{LN,RN,LF,RF}\}$}
\FOR{$k=1,\hdots, N_{val}$}
\STATE{Obtain network output $\textbf{p}_{[i]}(k)$ using Eq.~\eqref{eq:softmax_result}}
\IF{$k=1,\hdots, l_{s}$}
\STATE{$\textbf{p}_{[i]final}(k)=\textbf{p}_{[i]}(k)$}
\ELSE
\FOR{$j=k, \hdots, k-l_{s}$}
\STATE{Obtain $\bar{\textbf{p}}(k)$ using Eq.~\eqref{eq.stack_output}}
\ENDFOR
\STATE{Obtain $V_{x}(k-l_s),\hdots ,V_{x}(k-1)$}
\STATE{Calculate $\Omega_{k}(\bar{\textbf{V}}_{x}(k-1|k-l_{s}))$ using Eq.~\eqref{eq.weight}}
\STATE{Update final result $\textbf{p}_{[i]final}(k)$ using Eq.~\eqref{eq:geometric}}
\ENDIF
\ENDFOR
\end{algorithmic}
\end{algorithm}
\section{Experiment Results}
\label{sec:ExperimentResult}
\begin{table*}[]
\centering
\caption{The confusion matrix of validation data (LN Region)}
\label{tb:confusion_LN}
\begin{tabular}{ccc|ccccccccc|cc|}
\cline{4-14}
\multicolumn{3}{c|}{\multirow{3}{*}{}}                                                                           & \multicolumn{9}{c|}{\textbf{Actual}}                                                                                                                                                                                                                                                                   & \multicolumn{1}{c|}{\multirow{3}{*}{Total}} & \multirow{3}{*}{Precision (\%)} \\ \cline{4-12}
\multicolumn{3}{c|}{}                                                                                            & \multicolumn{4}{c|}{Dry}                                                                                                             & \multicolumn{4}{c|}{Wet}                                                                                                                & \multirow{2}{*}{Snow} & \multicolumn{1}{c|}{}                       &                                 \\ \cline{4-11}
\multicolumn{3}{c|}{}                                                                                            & \multicolumn{1}{c|}{Asphalt} & \multicolumn{1}{c|}{Cement}     & \multicolumn{1}{c|}{Gravel}     & \multicolumn{1}{c|}{Sand}         & \multicolumn{1}{c|}{Asphalt}    & \multicolumn{1}{c|}{Cement}     & \multicolumn{1}{c|}{Gravel}     & \multicolumn{1}{c|}{Sand}         &                       & \multicolumn{1}{c|}{}                       &                                 \\ \hline
\multicolumn{1}{|c|}{\multirow{9}{*}{\textbf{Prediction}}} & \multicolumn{1}{c|}{\multirow{4}{*}{Dry}} & Asphalt & \multicolumn{1}{c|}{6,906}   & \multicolumn{1}{c|}{4} & \multicolumn{1}{c|}{0} & \multicolumn{1}{c|}{24}  & \multicolumn{1}{c|}{\textbf{0}} & \multicolumn{1}{c|}{\textbf{4}} & \multicolumn{1}{c|}{\textbf{0}} & \multicolumn{1}{c|}{\textbf{0}}   & \textbf{8}            & \multicolumn{1}{c|}{6,946}                  & 99.4                            \\ \cline{3-14}
\multicolumn{1}{|c|}{}                                     & \multicolumn{1}{c|}{}                     & Cement  & \multicolumn{1}{c|}{6}       & \multicolumn{1}{c|}{6,673}      & \multicolumn{1}{c|}{7} & \multicolumn{1}{c|}{455} & \multicolumn{1}{c|}{\textbf{0}} & \multicolumn{1}{c|}{\textbf{1}} & \multicolumn{1}{c|}{\textbf{0}} & \multicolumn{1}{c|}{\textbf{0}}   & \textbf{10}           & \multicolumn{1}{c|}{7,152}                  & 93.3                            \\ \cline{3-14}
\multicolumn{1}{|c|}{}                                     & \multicolumn{1}{c|}{}                     & Gravel  & \multicolumn{1}{c|}{3}       & \multicolumn{1}{c|}{8}          & \multicolumn{1}{c|}{6,980}      & \multicolumn{1}{c|}{2}   & \multicolumn{1}{c|}{\textbf{0}} & \multicolumn{1}{c|}{\textbf{1}} & \multicolumn{1}{c|}{\textbf{0}} & \multicolumn{1}{c|}{\textbf{19}}  & \textbf{21}           & \multicolumn{1}{c|}{7,034}                  & 99.2                            \\ \cline{3-14}
\multicolumn{1}{|c|}{}                                     & \multicolumn{1}{c|}{}                     & Sand    & \multicolumn{1}{c|}{35}      & \multicolumn{1}{c|}{247}        & \multicolumn{1}{c|}{7}          & \multicolumn{1}{c|}{6,516}        & \multicolumn{1}{c|}{\textbf{0}} & \multicolumn{1}{c|}{\textbf{1}} & \multicolumn{1}{c|}{\textbf{0}} & \multicolumn{1}{c|}{\textbf{0}}   & \textbf{0}            & \multicolumn{1}{c|}{6,806}                  & 95.7                            \\ \cline{2-14}
\multicolumn{1}{|c|}{}                                     & \multicolumn{1}{c|}{\multirow{4}{*}{Wet}} & Asphalt & \multicolumn{1}{c|}{0}       & \multicolumn{1}{c|}{0}          & \multicolumn{1}{c|}{0}          & \multicolumn{1}{c|}{0}            & \multicolumn{1}{c|}{6,996}      & \multicolumn{1}{c|}{5} & \multicolumn{1}{c|}{0} & \multicolumn{1}{c|}{1}   & \textbf{0}            & \multicolumn{1}{c|}{7,002}                  & 99.9                            \\ \cline{3-14}
\multicolumn{1}{|c|}{}                                     & \multicolumn{1}{c|}{}                     & Cement  & \multicolumn{1}{c|}{16}      & \multicolumn{1}{c|}{2}          & \multicolumn{1}{c|}{0}          & \multicolumn{1}{c|}{0}            & \multicolumn{1}{c|}{4}          & \multicolumn{1}{c|}{6,919}      & \multicolumn{1}{c|}{1} & \multicolumn{1}{c|}{8}   & \textbf{4}            & \multicolumn{1}{c|}{6,954}                  & 99.5                            \\ \cline{3-14}
\multicolumn{1}{|c|}{}                                     & \multicolumn{1}{c|}{}                     & Gravel  & \multicolumn{1}{c|}{0}       & \multicolumn{1}{c|}{0}          & \multicolumn{1}{c|}{0}          & \multicolumn{1}{c|}{0}            & \multicolumn{1}{c|}{0}          & \multicolumn{1}{c|}{6}          & \multicolumn{1}{c|}{6,949}      & \multicolumn{1}{c|}{119} & \textbf{0}            & \multicolumn{1}{c|}{7,074}                  & 98.2                            \\ \cline{3-14}
\multicolumn{1}{|c|}{}                                     & \multicolumn{1}{c|}{}                     & Sand    & \multicolumn{1}{c|}{0}       & \multicolumn{1}{c|}{2}          & \multicolumn{1}{c|}{0}          & \multicolumn{1}{c|}{1}            & \multicolumn{1}{c|}{0}          & \multicolumn{1}{c|}{32}         & \multicolumn{1}{c|}{50}         & \multicolumn{1}{c|}{6,848}        & \textbf{8}            & \multicolumn{1}{c|}{6,941}                  & 98.7                            \\ \cline{2-14}
\multicolumn{1}{|c|}{}                                     & \multicolumn{2}{c|}{Snow}                           & \multicolumn{1}{c|}{34}      & \multicolumn{1}{c|}{64}         & \multicolumn{1}{c|}{6}          & \multicolumn{1}{c|}{2}            & \multicolumn{1}{c|}{0}          & \multicolumn{1}{c|}{31}         & \multicolumn{1}{c|}{0}          & \multicolumn{1}{c|}{5}            & 6,949                 & \multicolumn{1}{c|}{7,091}                  & 98.0                            \\ \hline
\multicolumn{3}{|c|}{Total}                                                                                      & \multicolumn{1}{c|}{7,000}   & \multicolumn{1}{c|}{7,000}      & \multicolumn{1}{c|}{7,000}      & \multicolumn{1}{c|}{7,000}        & \multicolumn{1}{c|}{7,000}      & \multicolumn{1}{c|}{7,000}      & \multicolumn{1}{c|}{7,000}      & \multicolumn{1}{c|}{7,000}        & 7,000                 & \multicolumn{2}{c|}{63,000}                                                   \\ \hline
\multicolumn{3}{|c|}{Recall (\%)}                                                                                & \multicolumn{1}{c|}{98.7}    & \multicolumn{1}{c|}{95.3}       & \multicolumn{1}{c|}{99.7}       & \multicolumn{1}{c|}{93.1}         & \multicolumn{1}{c|}{99.9}       & \multicolumn{1}{c|}{98.8}       & \multicolumn{1}{c|}{99.3}       & \multicolumn{1}{c|}{97.8}         & 99.3                  & \multicolumn{2}{c|}{\textbf{Accuracy (\%): 98.0}}                             \\ \hline
\end{tabular}
\end{table*}
%
\begin{table*}[]
\centering
\caption{The confusion matrix of validation data (RN Region)}
\label{tb:confusion_RN}
\begin{tabular}{ccc|ccccccccc|cc|}
\cline{4-14}
\multicolumn{3}{c|}{\multirow{3}{*}{}}                                                                           & \multicolumn{9}{c|}{\textbf{Actual}}                                                                                                                                                                                                                                                                   & \multicolumn{1}{c|}{\multirow{3}{*}{Total}} & \multirow{3}{*}{Precision (\%)} \\ \cline{4-12}
\multicolumn{3}{c|}{}                                                                                            & \multicolumn{4}{c|}{Dry}                                                                                                             & \multicolumn{4}{c|}{Wet}                                                                                                                & \multirow{2}{*}{Snow} & \multicolumn{1}{c|}{}                       &                                 \\ \cline{4-11}
\multicolumn{3}{c|}{}                                                                                            & \multicolumn{1}{c|}{Asphalt} & \multicolumn{1}{c|}{Cement}     & \multicolumn{1}{c|}{Gravel}     & \multicolumn{1}{c|}{Sand}         & \multicolumn{1}{c|}{Asphalt}    & \multicolumn{1}{c|}{Cement}     & \multicolumn{1}{c|}{Gravel}     & \multicolumn{1}{c|}{Sand}         &                       & \multicolumn{1}{c|}{}                       &                                 \\ \hline
\multicolumn{1}{|c|}{\multirow{9}{*}{\textbf{Prediction}}} & \multicolumn{1}{c|}{\multirow{4}{*}{Dry}} & Asphalt & \multicolumn{1}{c|}{6,977}   & \multicolumn{1}{c|}{4} & \multicolumn{1}{c|}{0} & \multicolumn{1}{c|}{10}  & \multicolumn{1}{c|}{\textbf{0}} & \multicolumn{1}{c|}{\textbf{7}} & \multicolumn{1}{c|}{\textbf{0}} & \multicolumn{1}{c|}{\textbf{1}}   & \textbf{9}            & \multicolumn{1}{c|}{7,008}                  & 99.6                            \\ \cline{3-14}
\multicolumn{1}{|c|}{}                                     & \multicolumn{1}{c|}{}                     & Cement  & \multicolumn{1}{c|}{0}       & \multicolumn{1}{c|}{6,761}      & \multicolumn{1}{c|}{4} & \multicolumn{1}{c|}{216} & \multicolumn{1}{c|}{\textbf{0}} & \multicolumn{1}{c|}{\textbf{1}} & \multicolumn{1}{c|}{\textbf{0}} & \multicolumn{1}{c|}{\textbf{0}}   & \textbf{60}           & \multicolumn{1}{c|}{7,042}                  & 96.0                            \\ \cline{3-14}
\multicolumn{1}{|c|}{}                                     & \multicolumn{1}{c|}{}                     & Gravel  & \multicolumn{1}{c|}{0}       & \multicolumn{1}{c|}{12}         & \multicolumn{1}{c|}{6,981}      & \multicolumn{1}{c|}{2}   & \multicolumn{1}{c|}{\textbf{0}} & \multicolumn{1}{c|}{\textbf{0}} & \multicolumn{1}{c|}{\textbf{0}} & \multicolumn{1}{c|}{\textbf{20}}  & \textbf{38}           & \multicolumn{1}{c|}{7,053}                  & 99.0                            \\ \cline{3-14}
\multicolumn{1}{|c|}{}                                     & \multicolumn{1}{c|}{}                     & Sand    & \multicolumn{1}{c|}{17}      & \multicolumn{1}{c|}{153}        & \multicolumn{1}{c|}{9}          & \multicolumn{1}{c|}{6,772}        & \multicolumn{1}{c|}{\textbf{0}} & \multicolumn{1}{c|}{\textbf{1}} & \multicolumn{1}{c|}{\textbf{0}} & \multicolumn{1}{c|}{\textbf{0}}   & \textbf{7}            & \multicolumn{1}{c|}{6,959}                  & 97.3                            \\ \cline{2-14}
\multicolumn{1}{|c|}{}                                     & \multicolumn{1}{c|}{\multirow{4}{*}{Wet}} & Asphalt & \multicolumn{1}{c|}{1}       & \multicolumn{1}{c|}{0}          & \multicolumn{1}{c|}{0}          & \multicolumn{1}{c|}{0}            & \multicolumn{1}{c|}{6,994}      & \multicolumn{1}{c|}{2} & \multicolumn{1}{c|}{0} & \multicolumn{1}{c|}{1}   & \textbf{0}            & \multicolumn{1}{c|}{6,998}                  & 99.9                            \\ \cline{3-14}
\multicolumn{1}{|c|}{}                                     & \multicolumn{1}{c|}{}                     & Cement  & \multicolumn{1}{c|}{1}       & \multicolumn{1}{c|}{0}          & \multicolumn{1}{c|}{0}          & \multicolumn{1}{c|}{0}            & \multicolumn{1}{c|}{6}          & \multicolumn{1}{c|}{6,960}      & \multicolumn{1}{c|}{2} & \multicolumn{1}{c|}{5}   & \textbf{3}            & \multicolumn{1}{c|}{6,977}                  & 99.8                            \\ \cline{3-14}
\multicolumn{1}{|c|}{}                                     & \multicolumn{1}{c|}{}                     & Gravel  & \multicolumn{1}{c|}{0}       & \multicolumn{1}{c|}{0}          & \multicolumn{1}{c|}{0}          & \multicolumn{1}{c|}{0}            & \multicolumn{1}{c|}{0}          & \multicolumn{1}{c|}{3}          & \multicolumn{1}{c|}{6,944}      & \multicolumn{1}{c|}{109} & \textbf{1}            & \multicolumn{1}{c|}{7,057}                  & 98.4                            \\ \cline{3-14}
\multicolumn{1}{|c|}{}                                     & \multicolumn{1}{c|}{}                     & Sand    & \multicolumn{1}{c|}{0}       & \multicolumn{1}{c|}{2}          & \multicolumn{1}{c|}{0}          & \multicolumn{1}{c|}{1}            & \multicolumn{1}{c|}{0}          & \multicolumn{1}{c|}{32}         & \multicolumn{1}{c|}{50}         & \multicolumn{1}{c|}{6,848}        & \textbf{8}            & \multicolumn{1}{c|}{6,941}                  & 98.7                            \\ \cline{2-14}
\multicolumn{1}{|c|}{}                                     & \multicolumn{2}{c|}{Snow}                           & \multicolumn{1}{c|}{4}       & \multicolumn{1}{c|}{70}         & \multicolumn{1}{c|}{4}          & \multicolumn{1}{c|}{0}            & \multicolumn{1}{c|}{0}          & \multicolumn{1}{c|}{3}          & \multicolumn{1}{c|}{0}          & \multicolumn{1}{c|}{4}            & 6,865                 & \multicolumn{1}{c|}{6,950}                  & 98.8                            \\ \hline
\multicolumn{3}{|c|}{Total}                                                                                      & \multicolumn{1}{c|}{7,000}   & \multicolumn{1}{c|}{7,000}      & \multicolumn{1}{c|}{7,000}      & \multicolumn{1}{c|}{7,000}        & \multicolumn{1}{c|}{7,000}      & \multicolumn{1}{c|}{7,000}      & \multicolumn{1}{c|}{7,000}      & \multicolumn{1}{c|}{7,000}        & 7,000                 & \multicolumn{2}{c|}{63,000}                                                   \\ \hline
\multicolumn{3}{|c|}{Recall (\%)}                                                                                & \multicolumn{1}{c|}{99.7}    & \multicolumn{1}{c|}{96.6}       & \multicolumn{1}{c|}{99.7}       & \multicolumn{1}{c|}{96.7}         & \multicolumn{1}{c|}{99.9}       & \multicolumn{1}{c|}{99.4}       & \multicolumn{1}{c|}{99.2}       & \multicolumn{1}{c|}{98.0}         & 98.0                  & \multicolumn{2}{c|}{\textbf{Accuracy (\%): 98.6}}                             \\ \hline
\end{tabular}
\end{table*}
\begin{table*}[]
\centering
\caption{The confusion matrix of validation data (LN Region w/o velocity feature)}
\label{tb:confusion_LN_no_vel}
\begin{tabular}{ccc|ccccccccc|cc|}
\cline{4-14}
\multicolumn{3}{c|}{\multirow{3}{*}{}}                                                                           & \multicolumn{9}{c|}{\textbf{Actual}}                                                                                                                                                                                                                                                                          & \multicolumn{1}{c|}{\multirow{3}{*}{Total}} & \multirow{3}{*}{Precision (\%)} \\ \cline{4-12}
\multicolumn{3}{c|}{}                                                                                            & \multicolumn{4}{c|}{Dry}                                                                                                               & \multicolumn{4}{c|}{Wet}                                                                                                                     & \multirow{2}{*}{Snow} & \multicolumn{1}{c|}{}                       &                                 \\ \cline{4-11}
\multicolumn{3}{c|}{}                                                                                            & \multicolumn{1}{c|}{Asphalt} & \multicolumn{1}{c|}{Cement}      & \multicolumn{1}{c|}{Gravel}      & \multicolumn{1}{c|}{Sand}         & \multicolumn{1}{c|}{Asphalt}    & \multicolumn{1}{c|}{Cement}       & \multicolumn{1}{c|}{Gravel}       & \multicolumn{1}{c|}{Sand}          &                       & \multicolumn{1}{c|}{}                       &                                 \\ \hline
\multicolumn{1}{|c|}{\multirow{9}{*}{\textbf{Prediction}}} & \multicolumn{1}{c|}{\multirow{4}{*}{Dry}} & Asphalt & \multicolumn{1}{c|}{4,793}   & \multicolumn{1}{c|}{54} & \multicolumn{1}{c|}{8}  & \multicolumn{1}{c|}{6}   & \multicolumn{1}{c|}{\textbf{0}} & \multicolumn{1}{c|}{\textbf{159}} & \multicolumn{1}{c|}{\textbf{37}}  & \multicolumn{1}{c|}{\textbf{174}}  & \textbf{222}          & \multicolumn{1}{c|}{5,453}                  & 87.9                            \\ \cline{3-14}
\multicolumn{1}{|c|}{}                                     & \multicolumn{1}{c|}{}                     & Cement  & \multicolumn{1}{c|}{661}     & \multicolumn{1}{c|}{5,271}       & \multicolumn{1}{c|}{67} & \multicolumn{1}{c|}{772} & \multicolumn{1}{c|}{\textbf{1}} & \multicolumn{1}{c|}{\textbf{1}}   & \multicolumn{1}{c|}{\textbf{0}}   & \multicolumn{1}{c|}{\textbf{9}}    & \textbf{45}           & \multicolumn{1}{c|}{6,827}                  & 77.2                            \\ \cline{3-14}
\multicolumn{1}{|c|}{}                                     & \multicolumn{1}{c|}{}                     & Gravel  & \multicolumn{1}{c|}{208}     & \multicolumn{1}{c|}{204}         & \multicolumn{1}{c|}{6,568}       & \multicolumn{1}{c|}{0}   & \multicolumn{1}{c|}{\textbf{1}} & \multicolumn{1}{c|}{\textbf{33}}  & \multicolumn{1}{c|}{\textbf{0}}   & \multicolumn{1}{c|}{\textbf{9}}    & \textbf{888}          & \multicolumn{1}{c|}{7,911}                  & 83.0                            \\ \cline{3-14}
\multicolumn{1}{|c|}{}                                     & \multicolumn{1}{c|}{}                     & Sand    & \multicolumn{1}{c|}{226}     & \multicolumn{1}{c|}{1,431}       & \multicolumn{1}{c|}{219}         & \multicolumn{1}{c|}{6,222}        & \multicolumn{1}{c|}{\textbf{0}} & \multicolumn{1}{c|}{\textbf{6}}   & \multicolumn{1}{c|}{\textbf{0}}   & \multicolumn{1}{c|}{\textbf{4}}    & \textbf{4}            & \multicolumn{1}{c|}{8,112}                  & 76.7                            \\ \cline{2-14}
\multicolumn{1}{|c|}{}                                     & \multicolumn{1}{c|}{\multirow{4}{*}{Wet}} & Asphalt & \multicolumn{1}{c|}{2}       & \multicolumn{1}{c|}{2}           & \multicolumn{1}{c|}{1}           & \multicolumn{1}{c|}{0}            & \multicolumn{1}{c|}{6,134}      & \multicolumn{1}{c|}{82}  & \multicolumn{1}{c|}{150} & \multicolumn{1}{c|}{191}  & \textbf{1}            & \multicolumn{1}{c|}{6,563}                  & 93.5                            \\ \cline{3-14}
\multicolumn{1}{|c|}{}                                     & \multicolumn{1}{c|}{}                     & Cement  & \multicolumn{1}{c|}{156}     & \multicolumn{1}{c|}{11}          & \multicolumn{1}{c|}{7}           & \multicolumn{1}{c|}{0}            & \multicolumn{1}{c|}{49}         & \multicolumn{1}{c|}{5,901}        & \multicolumn{1}{c|}{14}  & \multicolumn{1}{c|}{790}  & \textbf{54}           & \multicolumn{1}{c|}{6,982}                  & 84.5                            \\ \cline{3-14}
\multicolumn{1}{|c|}{}                                     & \multicolumn{1}{c|}{}                     & Gravel  & \multicolumn{1}{c|}{0}       & \multicolumn{1}{c|}{0}           & \multicolumn{1}{c|}{0}           & \multicolumn{1}{c|}{0}            & \multicolumn{1}{c|}{766}        & \multicolumn{1}{c|}{122}          & \multicolumn{1}{c|}{6,402}        & \multicolumn{1}{c|}{1043} & \textbf{1}            & \multicolumn{1}{c|}{8,334}                  & 76.8                            \\ \cline{3-14}
\multicolumn{1}{|c|}{}                                     & \multicolumn{1}{c|}{}                     & Sand    & \multicolumn{1}{c|}{20}      & \multicolumn{1}{c|}{1}           & \multicolumn{1}{c|}{3}           & \multicolumn{1}{c|}{0}            & \multicolumn{1}{c|}{49}         & \multicolumn{1}{c|}{563}          & \multicolumn{1}{c|}{391}          & \multicolumn{1}{c|}{4,690}         & \textbf{8}            & \multicolumn{1}{c|}{5,725}                  & 81.9                            \\ \cline{2-14}
\multicolumn{1}{|c|}{}                                     & \multicolumn{2}{c|}{Snow}                           & \multicolumn{1}{c|}{934}     & \multicolumn{1}{c|}{26}          & \multicolumn{1}{c|}{127}         & \multicolumn{1}{c|}{0}            & \multicolumn{1}{c|}{0}          & \multicolumn{1}{c|}{133}          & \multicolumn{1}{c|}{6}            & \multicolumn{1}{c|}{90}            & 5,777                 & \multicolumn{1}{c|}{7,093}                  & 81.4                            \\ \hline
\multicolumn{3}{|c|}{Total}                                                                                      & \multicolumn{1}{c|}{7,000}   & \multicolumn{1}{c|}{7,000}       & \multicolumn{1}{c|}{7,000}       & \multicolumn{1}{c|}{7,000}        & \multicolumn{1}{c|}{7,000}      & \multicolumn{1}{c|}{7,000}        & \multicolumn{1}{c|}{7,000}        & \multicolumn{1}{c|}{7,000}         & 7,000                 & \multicolumn{2}{c|}{63,000}                                                   \\ \hline
\multicolumn{3}{|c|}{Recall (\%)}                                                                                & \multicolumn{1}{c|}{68.5}    & \multicolumn{1}{c|}{75.3}        & \multicolumn{1}{c|}{93.8}        & \multicolumn{1}{c|}{88.9}         & \multicolumn{1}{c|}{87.6}       & \multicolumn{1}{c|}{84.3}         & \multicolumn{1}{c|}{91.5}         & \multicolumn{1}{c|}{67.0}          & 82.5                  & \multicolumn{2}{c|}{\textbf{Accuracy (\%): 82.2}}                             \\ \hline
\end{tabular}
\end{table*}
%
\begin{table*}[]
\centering
\caption{The confusion matrix of validation data (RN Region w/o velocity feature)}
\label{tb:confusion_RN_no_vel}
\begin{tabular}{ccc|ccccccccc|cc|}
\cline{4-14}
\multicolumn{3}{c|}{\multirow{3}{*}{}}                                                                           & \multicolumn{9}{c|}{\textbf{Actual}}                                                                                                                                                                                                                                                                            & \multicolumn{1}{c|}{\multirow{3}{*}{Total}} & \multirow{3}{*}{Precision (\%)} \\ \cline{4-12}
\multicolumn{3}{c|}{}                                                                                            & \multicolumn{4}{c|}{Dry}                                                                                                                & \multicolumn{4}{c|}{Wet}                                                                                                                      & \multirow{2}{*}{Snow} & \multicolumn{1}{c|}{}                       &                                 \\ \cline{4-11}
\multicolumn{3}{c|}{}                                                                                            & \multicolumn{1}{c|}{Asphalt} & \multicolumn{1}{c|}{Cement}      & \multicolumn{1}{c|}{Gravel}       & \multicolumn{1}{c|}{Sand}         & \multicolumn{1}{c|}{Asphalt}    & \multicolumn{1}{c|}{Cement}       & \multicolumn{1}{c|}{Gravel}       & \multicolumn{1}{c|}{Sand}           &                       & \multicolumn{1}{c|}{}                       &                                 \\ \hline
\multicolumn{1}{|c|}{\multirow{9}{*}{\textbf{Prediction}}} & \multicolumn{1}{c|}{\multirow{4}{*}{Dry}} & Asphalt & \multicolumn{1}{c|}{5,374}   & \multicolumn{1}{c|}{36} & \multicolumn{1}{c|}{4}   & \multicolumn{1}{c|}{0}   & \multicolumn{1}{c|}{\textbf{1}} & \multicolumn{1}{c|}{\textbf{57}}  & \multicolumn{1}{c|}{\textbf{140}} & \multicolumn{1}{c|}{\textbf{122}}   & \textbf{254}          & \multicolumn{1}{c|}{5,988}                  & 89.7                            \\ \cline{3-14}
\multicolumn{1}{|c|}{}                                     & \multicolumn{1}{c|}{}                     & Cement  & \multicolumn{1}{c|}{293}     & \multicolumn{1}{c|}{4,927}       & \multicolumn{1}{c|}{117} & \multicolumn{1}{c|}{157} & \multicolumn{1}{c|}{\textbf{0}} & \multicolumn{1}{c|}{\textbf{0}}   & \multicolumn{1}{c|}{\textbf{0}}   & \multicolumn{1}{c|}{\textbf{8}}     & \textbf{101}          & \multicolumn{1}{c|}{5,603}                  & 87.9                            \\ \cline{3-14}
\multicolumn{1}{|c|}{}                                     & \multicolumn{1}{c|}{}                     & Gravel  & \multicolumn{1}{c|}{363}     & \multicolumn{1}{c|}{579}         & \multicolumn{1}{c|}{6,568}        & \multicolumn{1}{c|}{0}   & \multicolumn{1}{c|}{\textbf{0}} & \multicolumn{1}{c|}{\textbf{12}}  & \multicolumn{1}{c|}{\textbf{0}}   & \multicolumn{1}{c|}{\textbf{27}}    & \textbf{895}          & \multicolumn{1}{c|}{8,444}                  & 77.8                            \\ \cline{3-14}
\multicolumn{1}{|c|}{}                                     & \multicolumn{1}{c|}{}                     & Sand    & \multicolumn{1}{c|}{157}     & \multicolumn{1}{c|}{1,428}       & \multicolumn{1}{c|}{209}          & \multicolumn{1}{c|}{6,843}        & \multicolumn{1}{c|}{\textbf{0}} & \multicolumn{1}{c|}{\textbf{0}}   & \multicolumn{1}{c|}{\textbf{0}}   & \multicolumn{1}{c|}{\textbf{6}}     & \textbf{55}           & \multicolumn{1}{c|}{8,698}                  & 78.7                            \\ \cline{2-14}
\multicolumn{1}{|c|}{}                                     & \multicolumn{1}{c|}{\multirow{4}{*}{Wet}} & Asphalt & \multicolumn{1}{c|}{1}       & \multicolumn{1}{c|}{1}           & \multicolumn{1}{c|}{0}            & \multicolumn{1}{c|}{0}            & \multicolumn{1}{c|}{6,367}      & \multicolumn{1}{c|}{309} & \multicolumn{1}{c|}{379} & \multicolumn{1}{c|}{163}   & \textbf{1}            & \multicolumn{1}{c|}{7,221}                  & 88.2                            \\ \cline{3-14}
\multicolumn{1}{|c|}{}                                     & \multicolumn{1}{c|}{}                     & Cement  & \multicolumn{1}{c|}{98}      & \multicolumn{1}{c|}{2}           & \multicolumn{1}{c|}{2}            & \multicolumn{1}{c|}{0}            & \multicolumn{1}{c|}{72}         & \multicolumn{1}{c|}{5,266}        & \multicolumn{1}{c|}{100} & \multicolumn{1}{c|}{1,339} & \textbf{46}           & \multicolumn{1}{c|}{6,925}                  & 76.0                            \\ \cline{3-14}
\multicolumn{1}{|c|}{}                                     & \multicolumn{1}{c|}{}                     & Gravel  & \multicolumn{1}{c|}{1}       & \multicolumn{1}{c|}{0}           & \multicolumn{1}{c|}{0}            & \multicolumn{1}{c|}{0}            & \multicolumn{1}{c|}{528}        & \multicolumn{1}{c|}{366}          & \multicolumn{1}{c|}{5,942}        & \multicolumn{1}{c|}{912}   & \textbf{0}            & \multicolumn{1}{c|}{7,749}                  & 76.7                            \\ \cline{3-14}
\multicolumn{1}{|c|}{}                                     & \multicolumn{1}{c|}{}                     & Sand    & \multicolumn{1}{c|}{40}      & \multicolumn{1}{c|}{0}           & \multicolumn{1}{c|}{9}            & \multicolumn{1}{c|}{0}            & \multicolumn{1}{c|}{32}         & \multicolumn{1}{c|}{959}          & \multicolumn{1}{c|}{429}          & \multicolumn{1}{c|}{4,333}          & \textbf{16}           & \multicolumn{1}{c|}{5,818}                  & 74.5                            \\ \cline{2-14}
\multicolumn{1}{|c|}{}                                     & \multicolumn{2}{c|}{Snow}                           & \multicolumn{1}{c|}{673}     & \multicolumn{1}{c|}{27}          & \multicolumn{1}{c|}{91}           & \multicolumn{1}{c|}{0}            & \multicolumn{1}{c|}{0}          & \multicolumn{1}{c|}{31}           & \multicolumn{1}{c|}{10}           & \multicolumn{1}{c|}{90}             & 5,632                 & \multicolumn{1}{c|}{6,554}                  & 85.9                            \\ \hline
\multicolumn{3}{|c|}{Total}                                                                                      & \multicolumn{1}{c|}{7,000}   & \multicolumn{1}{c|}{7,000}       & \multicolumn{1}{c|}{7,000}        & \multicolumn{1}{c|}{7,000}        & \multicolumn{1}{c|}{7,000}      & \multicolumn{1}{c|}{7,000}        & \multicolumn{1}{c|}{7,000}        & \multicolumn{1}{c|}{7,000}          & 7,000                 & \multicolumn{2}{c|}{63,000}                                                   \\ \hline
\multicolumn{3}{|c|}{Recall (\%)}                                                                                & \multicolumn{1}{c|}{76.8}    & \multicolumn{1}{c|}{70.4}        & \multicolumn{1}{c|}{93.8}         & \multicolumn{1}{c|}{97.8}         & \multicolumn{1}{c|}{91.0}       & \multicolumn{1}{c|}{75.2}         & \multicolumn{1}{c|}{84.9}         & \multicolumn{1}{c|}{61.9}           & 80.5                  & \multicolumn{2}{c|}{\textbf{Accuracy (\%): 81.4}}                             \\ \hline
\end{tabular}
\end{table*}

\subsection{Result of proposed method}
To validate the proposed algorithm, we utilized the validation dataset of Table \ref{tb:data}, which was not used for training. Datasets were collected from various places around the urban, rural, and mountain areas near Seoul city in Korea. Paved roads, such as asphalt and cement, were collected from the highway and local roads, including the proving ground of Korea Intelligent Automotive Parts Promotion Institute (KIAPI), and unpaved roads, such as gravel and sand, were collected in lane-free situations. Furthermore, additional data were acquired when it rained or snowed in each place to acquire data with different conditions on the road surfaces.
In Table~\ref{tb:confusion_LN} and Table~\ref{tb:confusion_RN}, the confusion matrices show the overall accuracy in bold at $98.0$\% and $98.6$\% for each region near the vehicle.
In addition, an essential value in the road classification system is the proportion of risk situations. For example, suppose a snow road is misclassified as dry or wet. In these cases, fatal car accidents can occur because of delivering incorrect friction data between the tire and the road to the control system. Hence, the risk situation is one of the most crucial issues and shows low values of about $0.01$\% for each region.
\subsection{Ablation study}
We conducted an ablation study to explain the reason for considering the vehicle speed in the feature vector. Since this paper aims to classify road surfaces while the vehicle moves at speeds within a certain range, it is necessary to use the selected feature from LiDAR considering speed to learn the classifier. Therefore, we evaluated the performance of the classifier for both cases. The performance of the classifier of the proposed method with the network input $\textbf{x}_{[i]}=[\bar{\textbf{s}}_{p}(k|k-l_{t})\,\bar{\textbf{s}}_{r}(k|k-l_{t})
\,\bar{\textbf{V}}_{x}(k|k-l_{t})]^{T}$ has an accuracy of 98.0\% and 98.6\% for LN region and RN region, respectively.
However, the performance of the classifier in a method that the vehicle speed is not selected as a feature vector, and the network input is $\textbf{x}_{[i]}=[\bar{\textbf{s}}_{p}(k|k-l_{t})\,
\bar{\textbf{s}}_{r}(k|k-l_{t})]^{T}$ drops to  82.2\% and 81.4\% accuracy for LN region and RN region, respectively as shown in Table~\ref{tb:confusion_LN_no_vel} and~\ref{tb:confusion_RN_no_vel}. In other words, to utilize spatiotemporal information, it is necessary to consider the vehicle's speed since the LiDAR feature is affected by vehicle motion during the vehicle moves. More accurate road surface classifiers can be conducted using spatiotemporal information through classifiers that take into account the distribution of features of LiDAR according to vehicle speed.
\subsection{Comparative study}
\begin{figure}[t]
\centering
\subfloat[][Structure of DEL]{\includegraphics[width=0.45\textwidth]{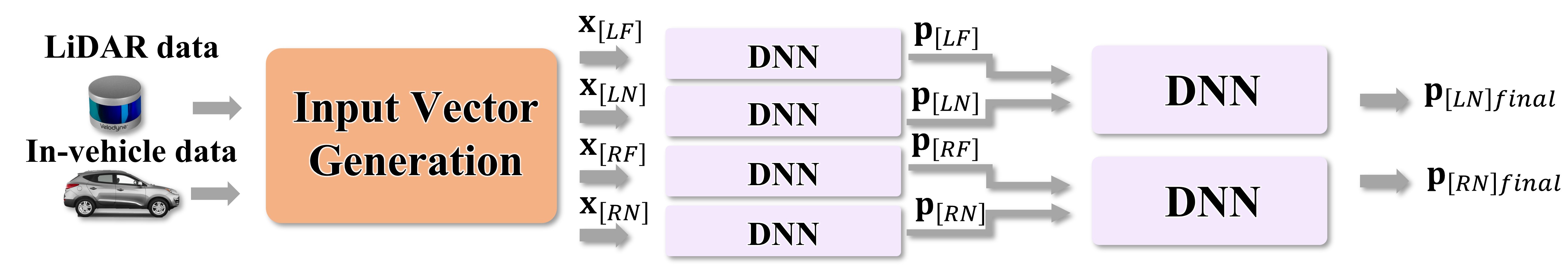}}
\hfil
\subfloat[][Structure of LSTM]{\includegraphics[width=0.45\textwidth]{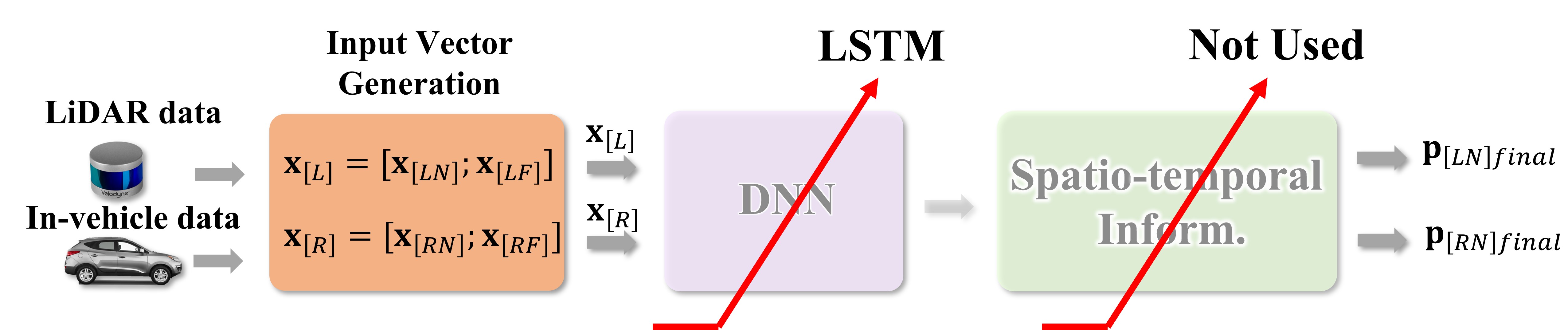}}
\hfil
\subfloat[][Structure of TWM]{\includegraphics[width=0.45\textwidth]{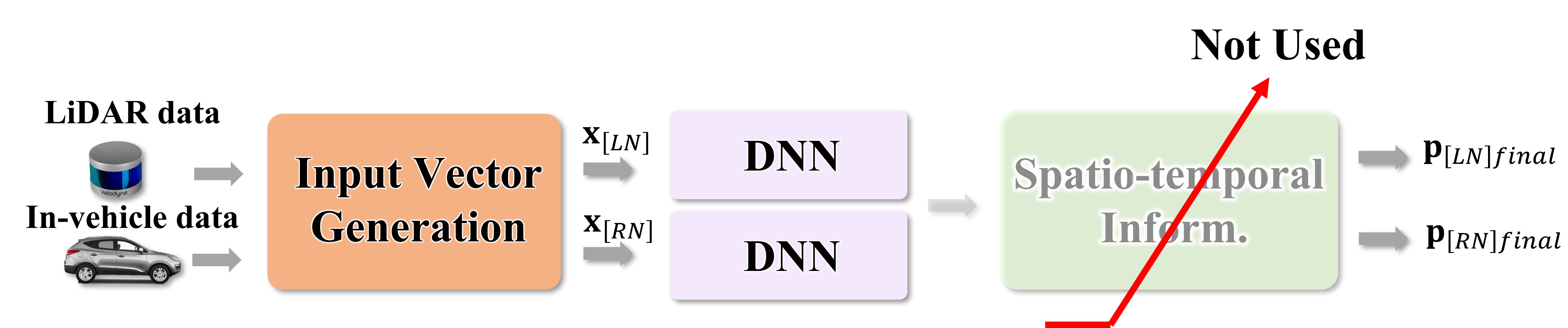}}
\hfil
\subfloat[][Structure of KNN]{\includegraphics[width=0.45\textwidth]{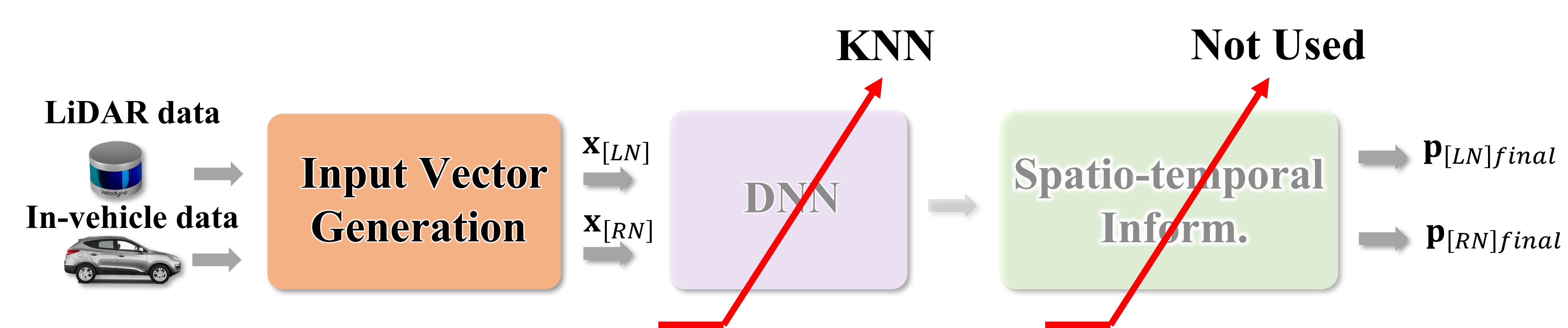}}
\hfil
\subfloat[][Structure of SVM]{\includegraphics[width=0.45\textwidth]{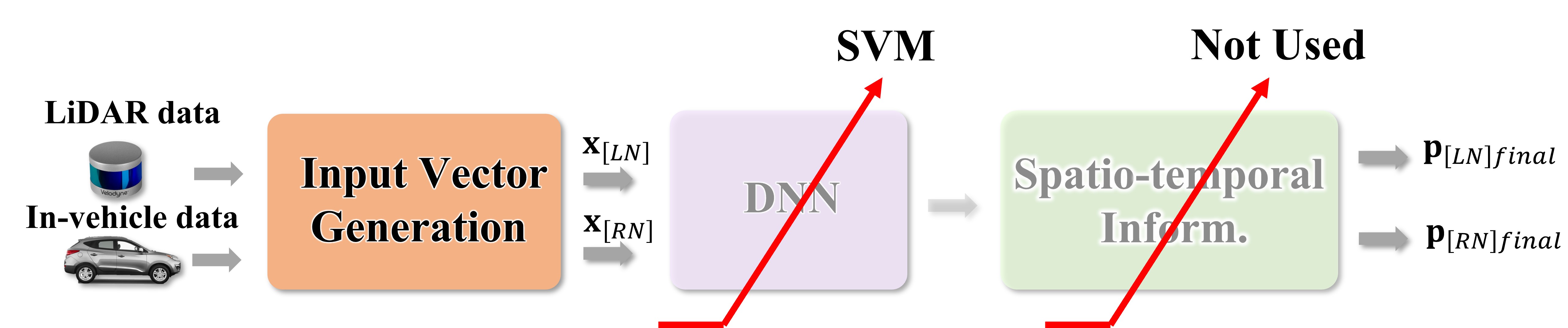}}
\caption{Simple architecture of DEL, LSTM, TWM, KNN, and SVM.}
\label{fig:other}
\end{figure}
We compared the proposed method with five other algorithms: Deep ensemble learning (DEL)~\cite{dietterich2000ensemble}, long-short term memory (LSTM)~\cite{Hochreiter1997lstm}, using only the time windowing method (TWM), K-nearest neighbor (KNN)~\cite{samet2007knn}, and support vector machine (SVM)~\cite{hearst1998svm}. A schematic structure of the five algorithms is shown in Fig.~\ref{fig:other}.
DEL uses spatiotemporal information similar to the proposed algorithm. However, there is a difference in which the parallel structure of DNN is used for data fusion, as shown in Fig~\ref{fig:other}~(a). The first DNN structure used for DEL is the same as the proposed method. The second DNN consists of three hidden layers, and the number of neurons in each layer is set to 100, 80, and 80, respectively. Both DNNs used non-linear functions, e.g., tangent-sigmoid functions, as transfer functions. LSTM is useful for learning time-sequence data. In this comparative study, the number of hidden units is 200 in LSTM. Each unit's state and gate activation functions are set to tangent-hyperbolic and sigmoid functions, respectively. The structure of LSTM is shown in Fig~\ref{fig:other}~(b). TWM excludes the spatiotemporal fusion method from the proposed method, as shown in Fig.~\ref{fig:other}~(c). The structure of the DNN used here is the same as the proposed method. KNN and SVM techniques have been utilized to classify the condition and type of road surface using LiDAR. Therefore, we compared using KNN and SVM, except for DNN and spatiotemporal fusion method in the proposed method as shown in Fig.~\ref{fig:other}~(d) and (e) to validate our algorithm.
In Table~\ref{tab:accuracy}, we compared the accuracy of SVM, KNN, TWM, DEL, and LSTM with the accuracy of the proposed method. It can be seen that the accuracy of the proposed method showed the highest values in both regions. Here, we can see that the DEL method has a similar performance to the proposed method because it contains spatiotemporal information. However, the proposed method can be analyzed to have better performance for moving ego vehicles since it uses a fusion method dependent on the ego vehicle's speed.
%
%
\begin{table}[t]
\centering
\caption{Accuracy of six methods used in comparative study(\%)}
\label{tab:accuracy}
\begin{tabular}{c|c|c|c|c|c|c|}
\cline{2-7}
\multirow{2}{*}{}                                         & \multirow{2}{*}{\textbf{Proposed}} & \multirow{2}{*}{\textbf{DEL}} & \multirow{2}{*}{\textbf{LSTM}} & \multirow{2}{*}{\textbf{TWM}} & \multirow{2}{*}{\textbf{KNN}} & \multirow{2}{*}{\textbf{SVM}} \\
                                                          &                                    &                               &                               &                               &                               &                                \\ \hline
\multicolumn{1}{|c|}{\multirow{2}{*}{\textbf{LN}}} & \multirow{2}{*}{\textbf{98.0}}     & \multirow{2}{*}{94.8}         & \multirow{2}{*}{87.3}         & \multirow{2}{*}{86.6}         & \multirow{2}{*}{85.7}         & \multirow{2}{*}{76.9}          \\
\multicolumn{1}{|c|}{}                                    &                                    &                               &                               &                               &                               &                                \\ \hline
\multicolumn{1}{|c|}{\multirow{2}{*}{\textbf{RN}}} & \multirow{2}{*}{\textbf{98.6}}     & \multirow{2}{*}{94.6}         & \multirow{2}{*}{87.0}         & \multirow{2}{*}{83.6}         & \multirow{2}{*}{83.5}         & \multirow{2}{*}{79.5}          \\
\multicolumn{1}{|c|}{}                                    &                                    &                               &                               &                               &                               &                                \\ \hline
\end{tabular}
\end{table}
%
\begin{figure}[t]
\centering
\subfloat[][Experimental environment]{\includegraphics[width=0.37\textwidth]{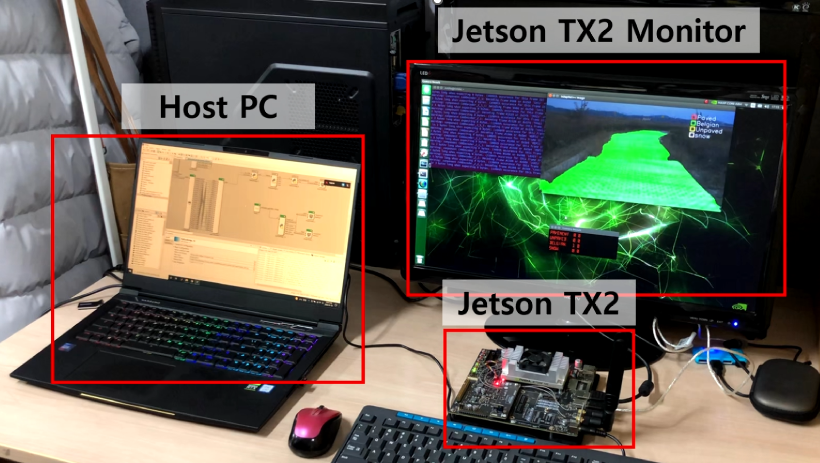}}
\hfil
\subfloat[][Overall structure of real time operation using jetson TX2 board]{\includegraphics[width=0.4\textwidth]{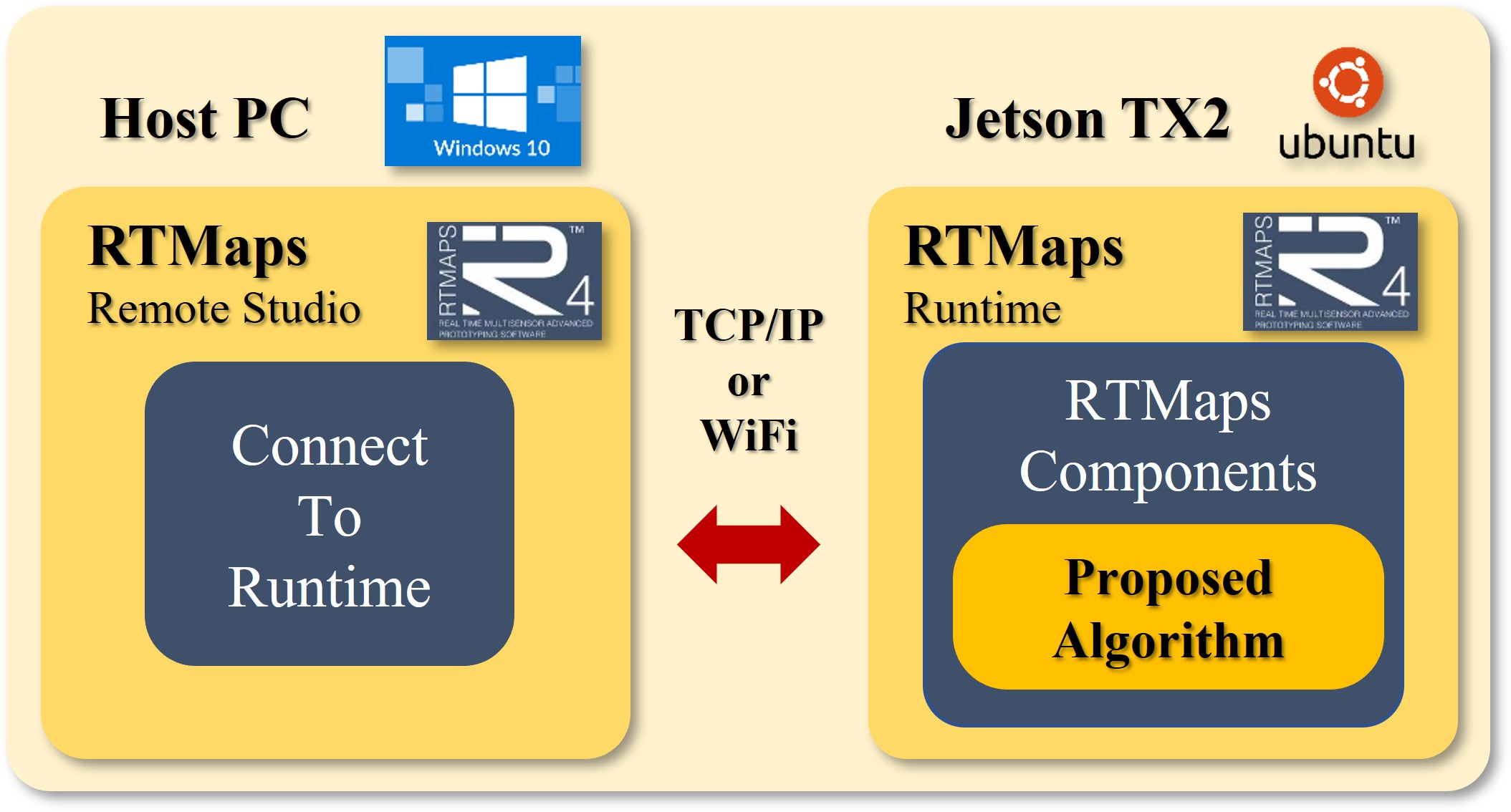}}
\caption{Experimental environment and overall structure of real time operation}
\label{fig:jetsonn}
\end{figure}
\subsection{Real time performance with Jetson TX2}
All the performances reported in the previous section were the outcome of a desktop personnel computer. To see the real time application of the proposed method,  we executed the proposed algorithm by porting it into a Jetson TX2 board from NVIDIA. The experimental environment is shown in Fig.~\ref{fig:jetsonn}~(a), and the overall structure of the experiment is shown in Fig.~\ref{fig:jetsonn}~(b). The Host PC was used only to provide an interface for RTMaps and was not involved in running the algorithm. The Host PC and the Jetson TX2 are communicated over TCP/IP protocol. The CPU of the Host PC is an  Intel i7-8700k six-core, and the CPU of the Jetson TX2 is Quad ARM A57 MPCore.

\begin{figure}[t]
\centering
\subfloat[][Classification probability of dry asphalt with Jetson TX2]{\includegraphics[width=0.45\textwidth]{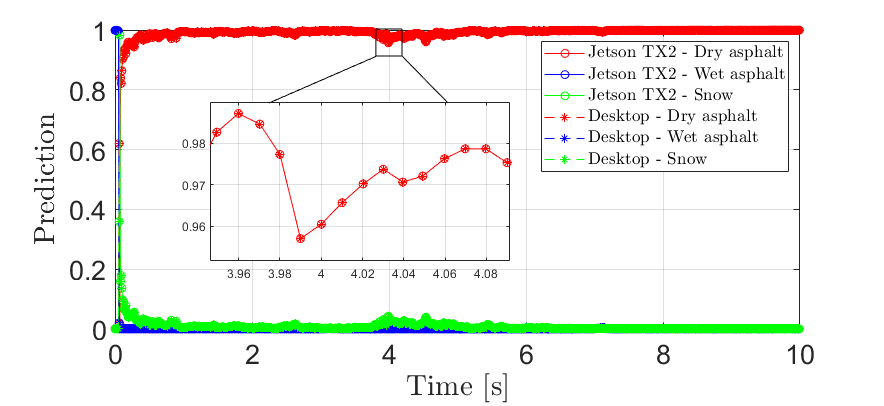}}
\hfil
\subfloat[][Computation time]{\includegraphics[width=0.45\textwidth]{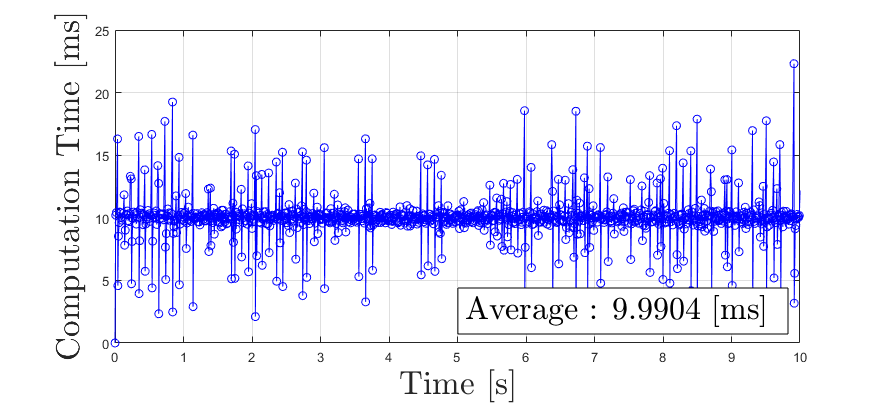}}
\caption{Experimental results using Jetson TX2}
\label{fig:jetson_result}
\end{figure}

We compared the results calculated in the desktop environment with those calculated in real time in the Jetson TX2. We showed an implementation example for dry asphalt, wet asphalt, and snow classes to see whether our algorithm runs in real time. Figure~\ref{fig:jetson_result}~(a) compares the results in desktop and Jetson TX2 for data with a ground truth of dry asphalt. The circle marker is the result of Jetson TX2, and the star marker is the result of the desktop. In the enlarged part, it can be seen that the results in the desktop and Jetson TX2 are almost the same.
%
%
In addition, we considered the computation time essential for the proposed algorithm to operate in real time, so we calculated the time when the algorithm was implemented in Jetson TX2. The computation time is shown in Fig.~\ref{fig:jetson_result}~(b), and its average is about 10 ms.

%

%
%
%
%
%
%
%
%
%
%
\section{CONCLUSIONS}
\label{sec:conclusion}
In this paper, we proposed a classification method for road surface conditions and types using LiDAR. We divided the front road area into four subregions to utilize spatial information. The reflectivity, the number of point clouds, and vehicle speed were selected as the feature vector from the LiDAR and in-vehicle sensors. Then, the DNN was used to classify road surface conditions and types for each subregion. The outputs of the DNN were used for spatiotemporal formulation to make the final classification results of the subregions near the ego vehicle. To validate the proposed algorithm, we compared it with five other methods. The accuracy of the subregions near the ego vehicle using the proposed method was the highest. In addition, we implemented the proposed algorithm into the Jetson TX2 board to show that our method can be applied in real time. The average computation time was about 10 ms, and the classification probability results were similar to desktop results. Therefore, the proposed method is expected to improve the safety of automated driving vehicles and be applicable in the real world.

\end{document}